\documentclass[11pt]{article}
\usepackage{graphicx}
\usepackage{epsfig}
\usepackage{amsfonts}
\usepackage{amssymb}
\usepackage{amsmath,amssymb}
\usepackage{url}
\usepackage{graphicx,epsf, epsfig, amssymb, amsmath, amsfonts, mathrsfs}
\usepackage{longtable}
\usepackage{bm}
\usepackage{color}
\usepackage{enumitem}
\usepackage{float}
\usepackage{caption}
\usepackage{subcaption}
\usepackage{upgreek}
\usepackage{cite}
\usepackage{mathtools}
\usepackage{titlesec}
\usepackage{lipsum}
\usepackage{xcolor}
\usepackage[section]{placeins}
\usepackage[colorlinks=true,linktocpage=true,linkcolor=blue,citecolor=blue,urlcolor=blue]{hyperref}
\usepackage{tensor}
\usepackage{multirow}
\usepackage[utf8]{inputenc}
\usepackage{authblk}
\usepackage{array}
\usepackage{breqn}

\newcommand{\beq}{\begin{equation}}
	\newcommand{\eeq}{\end{equation}}
\newcommand{\beqa}{\begin{eqnarray}}
	\newcommand{\eeqa}{\end{eqnarray}}

\newcommand{\dmoff}[1]{}

\usepackage{comment}

\begin{document}

\title{\Large{\bf 
Multipolar Proca stars: \\ electric, magnetic and hybrid solitons 
} }

\author{
{\large Carlos Herdeiro}$^{1}$,  
{\large Eugen Radu}$^{1}$, 
{\large Etevaldo dos Santos Costa Filho}$^{1}$ and
{\large Nicolas Sanchis-Gual}$^{2}$ 
\vspace*{0.2cm}
\\
$^{1}${\small Departamento de Matem\'atica da Universidade de Aveiro and } 
\\ {\small  Centre for Research and Development  in Mathematics and Applications (CIDMA),} 
\\ {\small    Campus de Santiago, 3810-193 Aveiro, Portugal}
\\ $^{2}${\small Departamento de Astronomía y Astrofísica, Universitat de València, Av. Vicent Andrés Estellés 19, 46100, Burjassot (Valencia), Spain}
}

\date{May 2026}

\maketitle

\begin{abstract}
We construct new families of everywhere regular, asymptotically flat solitons in the Einstein--Proca model, obtained as self-gravitating continuations of flat-spacetime (singular) Proca multipoles.
First we consider static and axially symmetric solutions, organized by a multipole number $\ell$. Two distinct classes arise: electric-type configurations, which include the spherical Proca stars as the $\ell=0$ case, and magnetic-type configurations, which have no spherical counterpart and start at $\ell=1$. Then we construct hybrid solutions as nonlinear superpositions of electric and magnetic multipoles. These have non-vanishing local angular momentum density but vanishing total angular momentum, and in some cases have no north-south $\mathbb{Z}_2$-symmetry.  
By performing dynamical evolutions of Proca stars in the new magnetic and hybrid sectors, we show they are unstable, decaying to the (static) prolate Proca stars or the (stationary) spinning Proca stars, previously identified as dynamically robust, electric sector configurations. In some cases, they can also collapse into a black hole.
\end{abstract}

\newpage

\tableofcontents

\section{Introduction}

Self-gravitating solitons supported by bosonic fields provide important examples of regular, horizonless solutions in general relativity, realizing to some extent Wheeler's vision of \textit{geons}~\cite{Wheeler:1955zz}. The best known case is that of scalar boson stars, which arise in the Einstein-Klein-Gordon model and have been extensively studied for several decades - see $e.g.$ the reviews~\cite{Jetzer:1991jr,Schunck:2003kk,Liebling:2012fv}. One decade ago, analogous solutions supported by massive vector fields - known as Proca stars - have been constructed in the Einstein-Proca model~\cite{Brito:2015pxa}.

 Proca stars share many qualitative features with scalar boson stars, including families of globally regular, asymptotically flat solutions characterized by a harmonic time dependence of the matter field and a conserved Noether charge. Since their discovery, several generalizations of Proca stars have been studied, including with vector self-interactions~\cite{Brihaye:2017inn,Minamitsuji:2018kof,Aoki:2022woy,Aoki:2022mdn}, electrically charged~\cite{SalazarLandea:2016bys,Mio:2025fnj}, with different asymptotics or topologies~\cite{Duarte:2016lig,Su:2023zhh,Li:2025bcz}, radially excited~\cite{Joaquin:2024quo}, static but non-spherical~\cite{Wang:2023tly,Herdeiro:2023wqf}, coupled to other fields~\cite{Herdeiro:2023lze,Cardoso:2021ehg,Jockel:2023rrm,Pombo:2023sih,Dzhunushaliev:2021vwn,Dzhunushaliev:2019ham,Herdeiro:2024pmv,Ma:2023vfa,Ma:2023bhb,Hernandez:2023tig}, in the  Newtonian limit~\cite{Jain:2021pnk,Nambo:2024hao}, with more than one Proca field~\cite{Zhang:2023rwc,Lazarte:2024jyr} or with non-minimal couplings~\cite{Babichev:2017rti,Minamitsuji:2017pdr,Brihaye:2021qvc}. On the other hand, applications of these solutions have been considered in the context of, $e.g.$, imaging~\cite{Cunha:2017wao,Herdeiro:2021lwl,Rosa:2022toh,Rosa:2022tfv}, orbital analysis~\cite{Delgado:2021jxd}, X-ray astrophysics~\cite{Shen:2016acv}, dynamics/stability~\cite{Sanchis-Gual:2017bhw,DiGiovanni:2018bvo,Sanchis-Gual:2018oui,Sanchis-Gual:2019ljs,Cunha:2022gde,Sanchis-Gual:2020mzb,Sanchis-Gual:2022mkk,Santos:2024vdm,Lazarte:2025wlw,Mio:2026wfh,Nambo:2025lnu}, tidal deformability~\cite{Herdeiro:2020kba}, universality relations~\cite{Adam:2024zqr}, or gravitational wave phenomenology~\cite{CalderonBustillo:2020fyi,CalderonBustillo:2020xms,CalderonBustillo:2022dph,Luna:2024kof,Palloni:2025mhn,Lorenzo-Medina:2025ylc}. Their emergence from beyond the standard model of particle physics has been discussed, $e.g.$,  in~\cite{Freitas:2021cfi}. Such growing literature on Proca stars over the last decade makes them a thriving research arena.

Vector fields introduce additional structure absent in the scalar case. In particular, the Proca field possesses multiple polarisation states and admits configurations with distinct angular structures. This richer field suggests new classes of solutions may be possible, even in the simplest Einstein-(complex) Proca model. Understanding the full space of such solutions is therefore an important step towards characterizing the soliton sector of this system.

A useful guide is provided by the corresponding solutions of the Proca equations in flat spacetime. For reference, recall first the scalar cousin model.
The Klein-Gordon equation 
on flat spacetime, in spherical coordinates, has solutions 
 $\Phi=e^{-i\omega t}\sum\limits_{\ell,m}{c_{\ell m}}R_\ell(r)Y_{\ell m}(\theta,\varphi)$,
  with the radial function 
\begin{equation}
R_\ell(r)=\frac{1}{\sqrt{r}}K_{\ell+\frac{1}{2}}\left(r\sqrt{\mu^2-\omega^2}\right) \  , 
\label{flat}
\end{equation}
where $\ell \in \mathbb{N}_0$; $K_j$ are  modified Bessel functions of the second kind;  $c_{\ell m}$ are constants; the harmonic time dependence has frequency $\omega \in \mathbb{R}^+$; $Y_{\ell m}$ are real spherical harmonics. The solutions~\eqref{flat} are  irregular, diverging at the origin; however, they are regularized by turning on gravity. 
Their non-linear self-gravitating versions, with a single $c_{\ell m}\neq 0$, yield multipolar stars~\cite{Herdeiro:2020kvf}.
Focusing on the axially symmetric sector  ($m=0$), one finds $e.g.$ monopolar ($\ell=0$), dipolar ($\ell=1$), quadrupolar ($\ell=2$), etc, scalar boson stars.
The ADM mass $M$ increases with $\ell$, fixing $\omega$. 
The ground state, thus, as in hydrogen, are the spherical states $(\ell=0)$.

Likewise, the Proca equation on flat spacetime admits multipolar configurations labeled by an integer $\ell$, analogous to electromagnetic multipoles. Again, these solutions are singular at the origin. This raises the question if gravity can regularize these singular multipolar configurations and turn them into everywhere regular solitons, as in the scalar case.

A first purpose of this work is to show that this is indeed the case. We construct new families of Proca stars that arise as self-gravitating continuations of flat-spacetime Proca multipoles. These solutions are static and axially symmetric, and are organized by a multipole number $\ell$. Two distinct classes occur. Electric-type configurations include the previously known spherical Proca stars~\cite{Brito:2015pxa}, corresponding to the $\ell=0$ case, the prolate Proca stars - which turn out to be the ground state of the model - as the $\ell=1$ case~\cite{Herdeiro:2023wqf}, as well as further higher multipolar generalizations. Magnetic-type configurations, on the other hand, are new and have no spherical counterpart, starting at $\ell=1$.

We also construct \textit{hybrid} generalizations of these solutions, mixing electric and magnetic multipoles. We find that some of these configurations have non-vanishing local angular momentum density but vanishing total angular momentum; others fail to have a $\mathbb{Z}_2$-north-south symmetry.

Our results reveal a richer structure of the Proca-star solution space than previously known, in the simplest Einstein-(complex)Proca model, including new magnetic and locally rotating sectors. For their physical viability a key question is their stability. The second purpose of this work is to show that all new magnetic and hybrid solutions evolved are actually \textit{unstable}, either decaying to the static prolate Proca stars or to the known rotating Proca stars~\cite{Herdeiro:2016tmi,Herdeiro:2017phl,Santos:2020pmh}. This underscores the insight that, albeit a larger landscape of solutions may be available, it may be quite restricted by dynamical stability.

The remainder of this paper is organized as follows. In section~\ref{section1} we present the Einstein-Proca model and the ansätze used for the fields in both the electric and magnetic cases. Section~\ref{solutions} discusses the static multipolar solutions. Section~\ref{sec:hybrid} presents the hybrid families. Section~\ref{sec:evolutions} deals with the dynamical evolutions and stability. We conclude in section~\ref{sec:remarks}.

\section{The model and the probe limit}
\label{section1}
We consider the action
\begin{equation}
\label{action}
\mathcal{S}= 
\int d^4x \sqrt{-g}\left[ 
\frac{1}{16 \pi  G}R
 -\frac{1}{4}\mathcal{F}_{\alpha\beta}\bar{\mathcal{F}}^{\alpha\beta}
-\frac{1}{2}\mu^2\mathcal{A}_\alpha\bar{\mathcal{A}}^\alpha\right] \ .
\end{equation}
with $R$
the Ricci scalar,
$\mathcal{A}$ a complex Proca field (a one-form) and $\mathcal{F}=d\mathcal{A}$.
 Also, $\mu$ is the mass of the field quanta,
while $g$ is the spacetime metric.

Varying  (\ref{action}) $w.r.t.$ the metric tensor and Proca potential 
 yields the equations of the model
\begin{eqnarray}
\label{EP}
E_{\alpha\beta}\equiv R_{\alpha\beta}-\frac{1}{2}R g_{\alpha\beta} - 8\pi G T_{\alpha \beta}=0,~~~
\nabla_\alpha\mathcal{F}^{\alpha\beta}=\mu^2 \mathcal{A}^\beta \ ,
\end{eqnarray}
with the energy-momentum tensor of the Proca field
\begin{eqnarray}
T_{\alpha\beta}=\frac{1}{2}
( \mathcal{F}_{\alpha \sigma }\bar{\mathcal{F}}_{\beta \gamma}
+\bar{\mathcal{F}}_{\alpha \sigma } \mathcal{F}_{\beta \gamma}
)g^{\sigma \gamma}
-\frac{1}{4}g_{\alpha\beta}\mathcal{F}_{\sigma\tau}\bar{\mathcal{F}}^{\sigma\tau}+\frac{1}{2}\mu^2\left[  
\mathcal{A}_{\alpha}\bar{\mathcal{A}}_{\beta}
+\bar{\mathcal{A}}_{\alpha}\mathcal{A}_{\beta}
-g_{\alpha\beta} \mathcal{A}_\sigma\bar{\mathcal{A}}^\sigma\right]\ . \ \ \ \ \ \ \  \ 
\label{procaemt}
\end{eqnarray}

Observe that the Proca equations \textit{completely} determine $ \mathcal{A}^\beta$ 
once $\mathcal{F}^{\alpha\beta}$ is known, while the electric-magnetic duality {\it is absent}. 
Moreover, unlike the Maxwell theory, the Proca model is not invariant under transformations of the potential 
$ \mathcal{A}\rightarrow  \mathcal{A}+\partial \chi$, where $\chi$ is a scalar function.
 In particular the Lorenz condition is implied by  the Proca equations
 (thus being a dynamical requirement, rather than a gauge choice):
\begin{equation}
\label{lorentz}
\nabla_\alpha\mathcal{A}^\alpha = 0 \ .
\end{equation}

The Proca action in (\ref{action})  possesses a global $U(1)$ symmetry, since it is invariant under the transformation 
$\mathcal{A}_\beta\rightarrow e^{i\chi}\mathcal{A}_\beta$, with $\chi$ constant; 
this implies the existence of a  4-current,  
\begin{equation}
\label{j}
j^\alpha=\frac{i}{2}\left[\bar{\mathcal{F}}^{\alpha \beta}\mathcal{A}_\beta-\mathcal{F}^{\alpha\beta}\bar{\mathcal{A}}_\beta\right] \ ,
\end{equation}
which is conserved by virtue of the field equations. 
Consequently, there exists a Noether charge, $Q$, 
obtained integrating the temporal component of the four-current on a space-like slice 
 $\Sigma$,
 \begin{equation} 
 Q=\int_\Sigma d^3x \sqrt{-g} j^t. 
 \label{q}
 \end{equation}

\subsection{ A static, axially symmetric Proca ansatz and flat spacetime  solutions}

Before discussing the solutions of the full Einstein-Proca system
(\ref{EP}),
it is useful to consider first the case of 
Proca equations 
on a (fixed) Minkowski spacetime background,
with the line element
\begin{equation}
\label{metricM}
ds^2= dr^2+r^2(d\theta^2+\sin^2 \theta d\varphi^2)-dt^2,
\end{equation}
where  $(r,\theta,\varphi)$ are  spherical coordinates with the usual range,
while $t$ is the time coordinate.
 
The most general Proca ansatz
$\mathcal{A}=\mathcal{A}_i dx^i$
leading to a $static$ axially symmetric
energy-momentum tensor
necessarily possesses no $\varphi$-dependence.
In what follows, 
we shall distinguish  two classes of solutions,
dubbed 'magnetic'  and  'electric',
depending on which Proca potential supports their existence. 
We remark that, in spite of the terminology used, 
both the electric and magnetic parts of the field strength are, in general, turned on  
 in both cases.

\subsubsection{The magnetic case}

This is the simplest case 
with a single potential\footnote{
The potential $H_3$ (and also
$H_1,H_2,V$ in the electric case, Eq. (\ref{ansatzE})) 
could be complex functions of $r$ and $\theta$;
however, in what follows we shall consider them real. 
}, 
\begin{eqnarray}
\label{ansatzM}
\mathcal{A} \equiv \mathcal{A}_m=
  i e^{ -i \omega t}  H_3(r,\theta) \sin \theta d\varphi ,
\end{eqnarray}
($\omega$ being the field frequency)
and the non-vanishing field strength components 
\begin{eqnarray}
\label{ansatzM-F}
\mathcal{F}_{r\varphi}=  i e^{ -i \omega t}  \partial_r H_3 \sin \theta,~~
\mathcal{F}_{ \theta\varphi}=  i e^{ -i \omega t}
 ( \sin \theta \partial_\theta H_3  +H_3 \cos \theta),~~
\mathcal{F}_{ \varphi t}= -\omega e^{ -i \omega t}  H_3 \sin \theta~.
\end{eqnarray}
An electric part of $\mathcal{F}$ is excited, due to
the $e^{ -i \omega t}$ factor in $\mathcal{A}_\varphi$.
In the absence of a $\varphi$-dependence in the ansatz,
 the Lorenz condition (\ref{lorentz})
is trivially satisfied.

\medskip

Given this ansatz,
the general solution of the Proca equations 
on the flat spacetime background (\ref{metricM})
 is a sum of
modes indexed by an integer $\ell$ and reads:
\begin{equation}
H_3(r,\theta)= \sum_{\ell \geq 1} \frac{d P_\ell(\cos \theta)}{d \theta} R_\ell(r) \ ,
\label{solM}
\end{equation}
where 
$P_\ell$ are Legendre polynomials ($\ell=0,1,\dots$).
The radial amplitude $R_\ell(r)$ has the following expression
\begin{equation}
\label{solMs}
R_\ell(r)=c_{1,\ell} \sqrt{r}I_{\frac{1}{2}(2\ell+1)}(r\sqrt{\mu^2- \omega^2})
+c_{2,\ell} \sqrt{r}K_{\frac{1}{2}(2\ell+1)}(r\sqrt{\mu^2- \omega^2}) \ ,
\end{equation} 
 with $c_{1,\ell}$, $c_{2,\ell} $ arbitrary constants; 
$I$ and $K$ are modified Bessel functions. 
As expected, the magnetic potential $H_3$ necessarily diverges at $r=0$
or at infinity.

The case $\ell=0$ is special; 
the generalization of the `Dirac monopole' solution in the Proca theory
exists for $\mu^2= \omega^2$ only, with
\begin{equation}
\mathcal{A}=Q_m e^{-i \omega t} \cos \theta d\varphi,
\end{equation}
where $Q_m$ is an arbitrary constant.
 However, this case is
 of no interest, since the singularities
 cannot be regularized
 even when including the gravity effects and/or a black hole horizon.

\subsubsection{The electric  case}

In this case, the temporal component of the Proca potential
is nonzero (hence the name 'electric').
However,
assuming again a harmonic time dependence of the field,
the Proca equations
imply that an
(axially symmetric) electric ansatz 
necessarily contains as well
two magnetic potentials $\mathcal{A}_r$ and $\mathcal{A}_\theta$, 
with 
\begin{eqnarray}
\label{ansatzE}
&&
\mathcal{A} \equiv 
\mathcal{A}_e=e^{ -i \omega t}\left(
 iV(r,\theta) dt  +\frac{H_1(r,\theta)}{r} dr
+H_2 (r,\theta)d\theta 
\right). 
\end{eqnarray}
The field strength tensor
containing also both electric and magnetic parts,
\begin{eqnarray}
\label{ansatzE-F}
\mathcal{F}_{r \theta}=    e^{ -i \omega t} 
\left( \partial_r H_2 -\frac{\partial_\theta  H_1}{r} \right),~~
\mathcal{F}_{ rt }=  i e^{ -i \omega t}
 \left( \frac{\omega H_1}{r} +\partial_r V \right),~~
\mathcal{F}_{ \theta t}= i e^{ -i \omega t} (\omega H_2+\partial_\theta  V)~.
\end{eqnarray}
The Lorenz condition  (\ref{lorentz})
results in the constraint
\begin{eqnarray}
\label{ansatzE-g}
 \partial_r (r H_1)+\frac{1}{\sin \theta} \partial_\theta  (\sin \theta H_2)
-r^2 \omega V=0.
\end{eqnarray}
As in the magnetic case,  
one considers a mode decomposition with
 \begin{equation}
 \label{solE} 
H_1(r, \theta)= \sum_{\ell \geq 0} h_1^{(\ell)} (r) P_\ell(\cos \theta),~~
H_2(r, \theta)= \sum_{\ell \geq 0} h_2^{(\ell)}(r) \frac{d P_\ell(\cos \theta)}{d \theta},~~
V(r, \theta)=  \sum_{\ell \geq 0}  v^{(\ell)}(r) P_\ell(\cos \theta)~,
\end{equation} 
with the Lorenz condition
(\ref{lorentz})
resulting in  
 \begin{eqnarray}
\label{lorentz1}
h_2^{(\ell)}=\frac{1}{\ell(\ell+1)}
 \left(
           r\frac{dh_1^{(\ell)}}{dr} +h_1^{(\ell)}-r^2 \omega v^{(\ell)} 
\right).
\end{eqnarray} 
 Then the remaining radial amplitudes solve the equations 
 \begin{eqnarray}
&&
\label{eqs-el}
 \frac{d^2v^{(\ell)}}{dr^2}
+\frac{2}{r }\frac{d v^{(\ell)}}{dr }
-\frac{ \ell(\ell+1)v^{(\ell)}}{r^2}
+(\omega^2-\mu^2)v^{(\ell)}=0,
\\ 
&&
\nonumber
 \frac{d^2 h_1^{(\ell)}}{dr^2}
+\frac{2}{r }\frac{dh_1^{(\ell)}}{dr }
-\left(\frac{(\ell(\ell+1) }{r^2} 
+\mu^2-\omega^2 
\right)
h_1^{(\ell)}=2\omega v^{(\ell)}.
\end{eqnarray}
Unfortunately,
a simple  solution can be written 
 for the electrostatic potential only\footnote{The Eq.  for $h_1^{(\ell)}$
can also be solved, being the sum of a part of the form (\ref{solE1}),
plus a very complicated expression (too long to include here), 
originating in the contribution of the $v^{(\ell)}$-term.
}, with
\begin{equation}
\label{solE1}
v^{(\ell)}(r)=c_{1,\ell} \frac{1}{\sqrt{r}}I_{\frac{1}{2}(2\ell+1)}(r\sqrt{\mu^2-\omega^2})
+c_{2,\ell} \frac{1}{\sqrt{r}} K_{\frac{1}{2}(2\ell+1)}(r\sqrt{\mu^2-\omega^2}),
\end{equation}
(where $c_{1,\ell}$, $c_{2,\ell}$ 
are integration constants)
which is again divergent at $r=0$ or at $r\to \infty$.

\section{Einstein-Proca static solutions }
\label{solutions}

It is natural to expect that,
analogous to the scalar field case \cite{Herdeiro:2020kvf}, 
the inclusion of gravity 
would regularize the $r=0$ singularity 
 of the flat spacetime Proca solutions, 
 while preserving the  exponential decay at infinity. 
The resulting solitonic Einstein-Proca configurations 
could be viewed as
 the non-linear continuation of the individual $\ell$-modes 
which enter the flat spacetime solution. 
However, in the presence of gravity, the labeling 
of solutions
in terms of $\ell$ is not
fully unambiguous, since, for a given parity  ${\cal P}$
(see below),
the Proca potentials are a superposition\footnote{
Moreover,   families of PS 
solutions with different $\ell$ (and the same parity)
can merge, see the example below.
}
of all $\ell$-angular modes with the same ${\cal P}$
(with different radial amplitudes).
Still, for a given $\ell$,
the corresponding angular mode in the Proca Ansatz 
 (\ref{solM}), 
(\ref{solE})
gives the dominant contribution in the far field,
at least for values of 
 $\omega$ close to the maximal one.

\medskip

In the absence of analytic methods to tackle the fully non-linear Einstein-Proca
system,
 we shall resort to numerical methods to construct these solutions, 
by solving a boundary value problem.
Following the  terminology in the flat spacetime case, we
shall consider
two  separate cases,
corresponding to
$magnetic$
and
$electric$   solutions,
distinguished by the employed Proca ansatz, which is  
(\ref{ansatzM})
or
(\ref{ansatzE}), respectively.

\subsection{The $\ell=0$ case: spherically symmetric 
$electric$
Proca Stars}
\label{sph}
 
\begin{figure}[h!]
\begin{center}
 \includegraphics[height=.34\textwidth, angle =0 ]{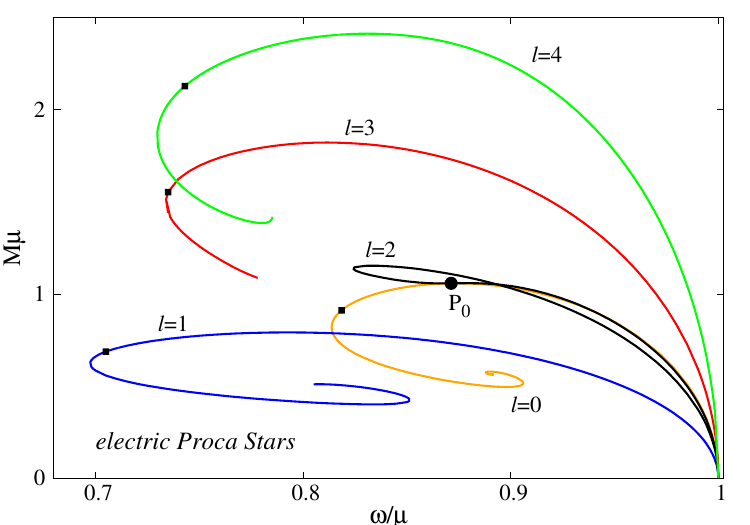} 
 \includegraphics[height=.34\textwidth, angle =0 ]{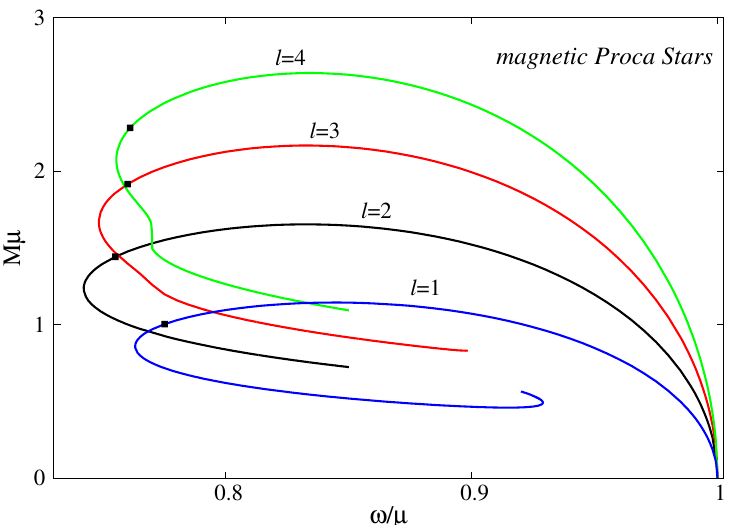}
\caption{ 
{\small
The frequency-mass diagram is shown for the first electric (left panel) and magnetic (right panel) families of 
 solitonic solutions in Einstein-Proca theory.
The Proca stars between the maximal frequency and configurations
marked with a (square) black dot have $M<\mu Q$. 
}
}
\label{wM}
\end{center}
\end{figure}

Before discussing the more general case, 
it is useful to briefly review the 
static, spherically symmetric Proca star solutions.
The Proca ansatz here corresponds to the case
$\ell=0$
in the general
electric ansatz (\ref{ansatzE}).
Furthermore,
to make contact with the results in Ref.
\cite{Brito:2015pxa},  
we take
 \begin{eqnarray}
\label{ansatz-sph1}
H_1=i r g(r),~~H_2=0,~~V=-i f(r),
\end{eqnarray} 
and use  
a Schwarzschild-like  metric ansatz
\begin{eqnarray}
	\label{ansatz-sph2}
ds^2= \frac{dr^2}{N(r)} +r^2 (d\theta^2+ \sin^2\theta d\varphi^2)-N(r)\sigma^2(r)dt^2,~~{\rm with}~~ N(r)=1-\frac{2m(r)}{r} ,
\end{eqnarray}
 $m(r),\sigma(r)$ being the (local) mass and the redshift function, respectively.
The equations satisfied by the functions $m , \sigma$ and $f , g $
take a simpler form for this Ansatz, with
\begin{eqnarray}
\label{eq-sph1}
&&
m' = 4 \pi G   r^2 \left[ \frac{(f'-\omega g)^2}{2 \sigma^2} + \frac{\mu^2}{2} \left(g^2 N + \frac{f^2}{N \sigma^2} \right) \right] \ , 
~~
\frac{\sigma'}{\sigma}  = 4 \pi G  r \mu^2 \left(g^2 + \frac{f^2}{\sigma^2 N^2} \right) \ , 
\\
&&
\omega g - f'  = \frac{\mu^2}{\omega} \sigma^2 N g \ ,~~
 \frac{d}{dr} \left[ \frac{r^2 (f'-\omega g)}{\sigma} \right] =\frac{\mu^2 r^2 f}{\sigma N}  \ .
\end{eqnarray}
where the prime denotes differentiation with respect to $r$.
The quantities of interest are the ADM mass $M$,
 which can be read from
 the asymptotic value of 
the function $m(r)$, 
and the Noether charge 
$Q=\frac{4\pi \mu^2}{\omega}\int^{\infty}_0dr\, r^2g(r)^2\sigma(r)N(r)$.
The profiles of  typical solutions
were shown in Ref. \cite{Brito:2015pxa},
 together with a discussion of their asymptotics and
main properties\footnote{
In this work  
we shall restrict to the case of
 the fundamental  solutions reported in 
Ref. \cite{Brito:2015pxa},
which contain a single node for $f$ and no nodes for $g$.
However, spherical {\it excited} solutions exist as well~\cite{Joaquin:2024quo}.
}.
For example,  
a subset of solutions
is perturbatively stable $w.r.t.$
spherically symmetric perturbations~\cite{Santos:2024vdm}.
The $\ell=0$ Proca stars
exist for a limited range of frequencies,
and there is a maximal mass and Noether charge
attained for an intermediate frequency,
see the corresponding $\ell=0$ curve in
Figure \ref{wM} (left).

\subsection{The axially symmetric case: boundary conditions
and quantities of interest}

Returning to the axially symmetric case (and an arbitrary $\ell$),
we shall use a 
 static, axially symmetric  metric ansatz 
with three unknown functions $F_i$:
\begin{eqnarray}
	\label{metric}
	ds^2= -e^{2F_0(r,\theta)}   dt^2+ e^{2F_1(r,\theta)}(dr^2+ r^2  d\theta^2)
	+e^{2F_2(r,\theta)}r^2 \sin^2\theta d\varphi^2.
\end{eqnarray}  
Then the equations (\ref{EP})
 reduce  to solving
three Einstein equations 
together with  
 one (three) Proca equations for the magnetic (electric) cases, respectively.

Although 
these cases are approached separately,
the  boundary conditions (BCs) satisfied by the metric functions
are the same  (with $i=0,1,2$):
\begin{eqnarray} 
\label{bc-metric}
\partial_r F_i\big|_{r=0} =  0,~~ F_i\big|_{r=\infty}  =0, ~~ \partial_\theta  F_i\big|_{\theta =0,\pi }=0 ~. 
\end{eqnarray}
In addition, the absence of conical singularities imposes 
that $F_1=F_2$ should hold at $\theta =0,\pi$,
a condition which is verified from the numerical output.

The BCs satisfied by the Proca magnetic potential
are 
\begin{eqnarray} 
\label{bc-mag1}
 H_3 |_{r=0}=0 \ , \  H_3|_{r=\infty} =0,~~~ H_3|_{\theta =0,\pi } =0 ,
\end{eqnarray}
while
in the electric case we impose 
\begin{eqnarray} 
\label{bc-el} 
&&
 H_1 |_{r=0}= H_2 |_{r=0}= \partial_r V |_{r=0}=0 \ , 
~
 H_1|_{r=\infty}= H_2|_{r=\infty}=V|_{r=\infty} =0 ,~ 
\\
\nonumber
&&
\partial_\theta H_1 |_{\theta =0,\pi }= H_2 |_{\theta =0,\pi }  
=\partial_\theta V |_{\theta =0,\pi }= 0 ,~{~}
\end{eqnarray}
except for odd value of $\ell$,
where  we require $V |_{r=0}=0$.

Both the electric and magnetic
 configurations 
considered in this work
naturally
split into two classes, distinguished by their behaviour $w.r.t.$
a reflection along the equatorial plane $\theta=\pi/2$.
While the geometry is left invariant under this transformation,
$F_i(\theta)=F_i(\pi-\theta)$ ($i=0,1,2$),
the situation with the Proca potentials
is more complicated.
Depending on the value of the solution index $\ell$,
one can distinguish naturally between ${\cal P}=1$
(even parity, no change of sign when taking $\pi \to \pi-\theta$)  
and ${\cal P}=-1$ (odd parity) cases.
In the magnetic case, the parity for $H_3$ is
${\cal P}=(-1)^{\ell+1}$.
For electric configurations,
the situation is not uniform along the different ansatz functions, and is summarized in Table 1.
 ${\cal P}= 1$ implies a Neumann  boundary condition at $\theta=\pi/2$
 for the corresponding function, 
 while the
  Dirichlet condition holds for ${\cal P}= -1$.
  
This choice of parity assures that
 the components of the energy-momentum tensor are 
invariant under a reflection along the equatorial plane. 

\begin{table}[h!]
	\begin{center}
		\label{tab:table1}
		\begin{tabular}
			{  c || c | c | c ||c ||}
		        $~$	  & $H_1$ & $H_2$ & V& $H_3$
		\\
			\hline
		${\cal P}$ &	$(-1)^{\ell}$ &  $(-1)^{\ell+1}$ & $(-1)^{\ell}$ & $(-1)^{\ell+1}$  \\ 
		\end{tabular}
	\end{center}
		\caption{The parity of the Proca field potentials
        of static configurations
        with a given index ${\ell}$.}
\end{table}

One can construct an approximate solution 
compatible with the above BCs;
in particular, the large-$r$ form of the Proca potentials
reveals that the bound state condition
$\omega \leq \mu$
found for other localized solutions of the Einstein-Proca system
still holds in this case.  

\medskip

The solutions possess a mass $M$
and a  
quadrupole mass-moment $M_2$,
which 
can be read off from the asymptotic expansion of the metric 
component
$g_{tt}$ 
\cite{Pappas:2012ns},  
\begin{eqnarray}
\label{asym1}
&&
g_{tt}= -1
+\frac{2M}{r}
-2\left(\frac{M}{r}\right)^2
+\left( \frac{4}{3}\left[1-\frac{B_0}{2M^2}\right] -\frac{2\nu_2}{M^3} P_2(\cos\theta) 
\right)
\left(\frac{M}{r}\right)^3+\dots,~~
\\
\label{M2}
&&
{~~~~~~~ }{\rm with}~~
M_2= -\nu_2-\frac{4}{3}\left(\frac{1}{4}+\frac{B_0}{M^2}\right)M^3~.
\end{eqnarray}
Here, $B_0$ and $\nu_2$ are constant coefficients.

Another quantity of interest is the Noether charge,
 which is given by
 \begin{eqnarray} 
&&
\nonumber
{\rm electric~case:}~~~~~
 \label{Qe} 
Q =2\pi \int_0^\infty dr \int_0^\pi d\theta 
~
e^{F2-F0}\sin \theta
\bigg[
H_1(r\partial_r V+\omega H_1)
+H_2  ( \partial_\theta V+\omega H_2) 
\bigg] ,~~
\\
&&
 \label{Qm} 
{\rm magnetic~case:} ~~~
Q =2\pi ~\omega \int_0^\infty dr \int_0^\pi d\theta 
~
e^{2F_1-F2-F0}\sin \theta
H_3^2.
 \end{eqnarray}

In numerics  
we set $\mu = 1$, $4 \pi G = 1$,
by using a scaled radial coordinate
$r \to \mu r$
and scaled Proca potentials
$(H_i,V) \to \sqrt{4 \pi G}(H_i,V)$.
The Einstein-Proca equations 
are solved subject to the above boundary conditions,
by using a professional elliptic PDE solver \cite{schoen}
based on the Newton-Raphson procedure, employing a
compactified radial coordinate
$x=r/(r+c)$,
with $c$ a properly chosen constant,
usually taken to be one.
The Einstein equations 
$E_{r}^r-E_\theta^\theta$
and 
$E_{r}^\theta$
are not solved directly, being treated as constraints
and  
 providing an indication of the numerical accuracy.
In the electric case we have monitored the  Lorenz condition 
(\ref{lorentz}) 
as a further test of numerics.
The numerical error for the solutions
reported in this work is estimated to be typically
$ \lesssim 10^{-4}$.

In all cases (except for the electric $\ell=0$ one),
the initial guess in the numerics was the corresponding flat space $\ell$-mode
which vanishes as $r \to \infty$ (with a frequency close to the maximal one), while the divergence at $r=0$
was regularized via  a cutoff procedure.

\subsection{Properties of the solutions}

 Let us start by exhibiting in Figure \ref{wM}  the \{frequency-mass\}-diagram for
the first few values of $\ell$ 
and both the electric and the magnetic cases. The electric case has already been shown - for the first three curves - in~\cite{Herdeiro:2023wqf}. It has the striking feature that, for large frequencies, the lowest mass curve is the $\ell=1$, the true ground state of the model. In this respect, the magnetic solutions follow the more standard (hydrogen-like) mass hierarchy, with the mass increasing with $\ell$, as in the case of the scalar multipolar stars~\cite{Herdeiro:2020kvf}.

For all curves,  as $\omega \to \mu$, 
the Proca stars become spatially diluted,
while their mass vanishes. 
Reducing $\omega$ from its maximal value,
  $M$ increases,
	approaching a maximum $M_{max}$ for some intermediate value
  $\omega_M$.
	As the frequency further decreases a minimal value $\omega_{min}$ is reached, with a backbending towards a secondary branch of solutions backwards in $\omega$.
It is likely that the $M(\omega)$ curve 
 follows a spiral 
towards a central critical configuration with $\omega_{c}$. 
However, this behaviour is rather difficult to follow, since 
in general
the numerical
accuracy starts to decrease towards the end of the 2nd branch. We remark that $\{ \omega_M,\omega_{min},\omega_{c} \}$ 
and $M_{max}$
depend on the families of considered Proca stars.

A similar picture is found for the corresponding Noether charge curves 
$Q(\omega)$.
For all curves, 
$M < \mu Q$
 from $\omega = \mu$ down to almost the
minimal allowed frequency $\omega_{min}$,
the corresponding configurations being marked  with
black dots in Figure \ref{wM}. In this region, the Proca stars have less energy than the same number of free Proca particles. In other words, they are gravitationally bound.

A slightly odd case with respect to the above pattern is provided by the 
electric  $\ell=2$ solutions.
As seen in Figure   \ref{wM}  
(left panel)
the  $\ell=2$   curve intersects
 the $\ell=0$ curve
for a critical frequency
$\omega_{critical} \simeq 0.87129 \mu.$
As this configuration 
(marked as $P_0$ in the plots) 
is approached, the magnetic 
function $H_2$ trivializes, 
while
$(V,H_1)$
and the metric functions $F_i$
possess no angular dependence (with $F_1=F_2$),
$i.e.$ the spherical symmetry is recovered. It is
interesting to remark that the $M(\omega)$-curve
for the $\ell=2$ solutions
further continues back to the maximal frequency.

 This behaviour is related 
 to the existence of a smooth  
 $\ell=2$ 
 axial perturbation
 of the spherically symmetric Proca stars,
 with a normalizable 'zero mode'.
 As usual,
 this indicates a bifurcation towards a new branch of solutions, as discussed in
 \cite{Liang:2025myf}.

 \medskip
 
\begin{figure}[h!]
	\setlength{\unitlength}{1cm}
	\begin{picture}(15,12)
		\put(-1.,0){\epsfig{file=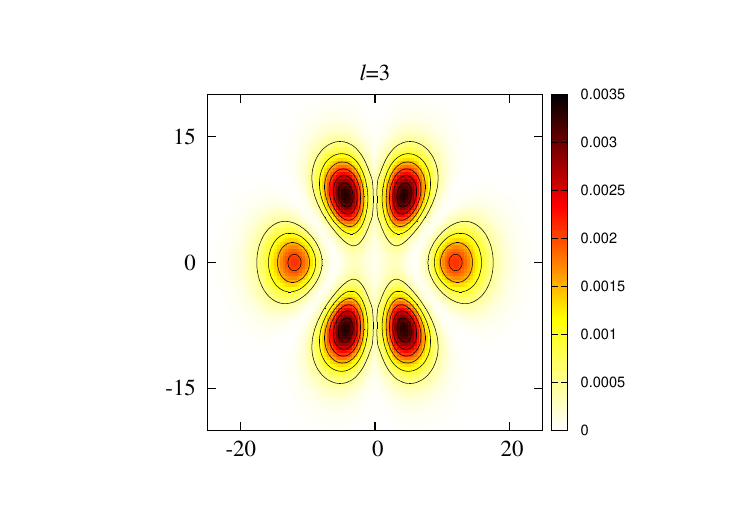,width=10cm}}
		\put(7, 0. ){\epsfig{file=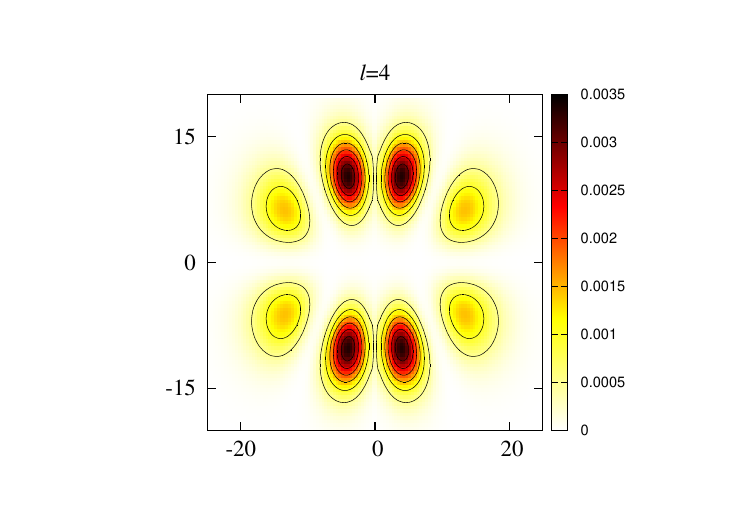,width=10cm}}
		\put(-1,6){\epsfig{file=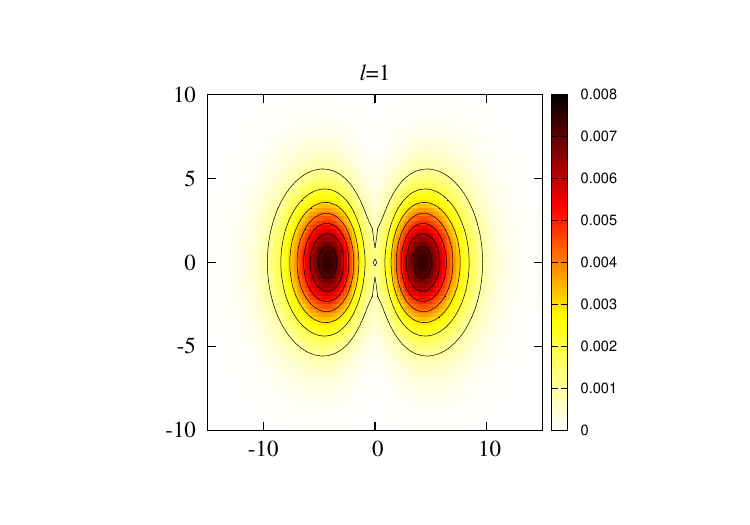,width=10cm}}
		\put(7,6){\epsfig{file=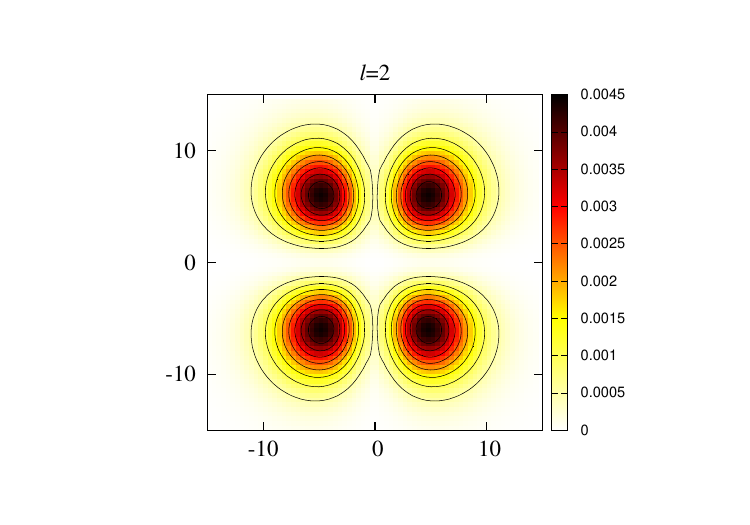,width=10cm}}
	\end{picture}
	\caption{
		The energy density is shown in $(x,z)$-plane for
		typical magnetic solutions
		with $\ell=1,2,3,4$
		(with $x=r \sin \theta \cos \varphi$,
		$z=r \cos \theta$; note that the solutions possess an azimuthal 
		symmetry, with no $\varphi$-dependence.)
	}
	\label{En-magnetic}
\end{figure}

\begin{figure}[h!]
	\setlength{\unitlength}{1cm}
	\begin{picture}(15,12)
		\put(-1.,0){\epsfig{file=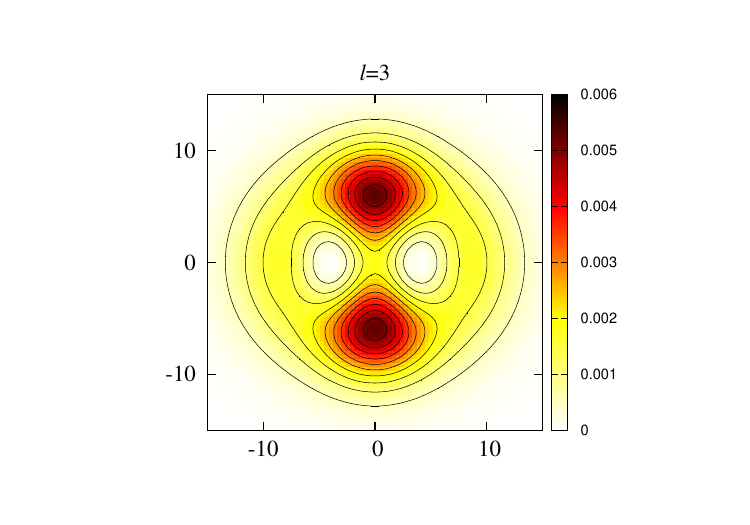,width=10cm}}
		\put(7, 0. ){\epsfig{file=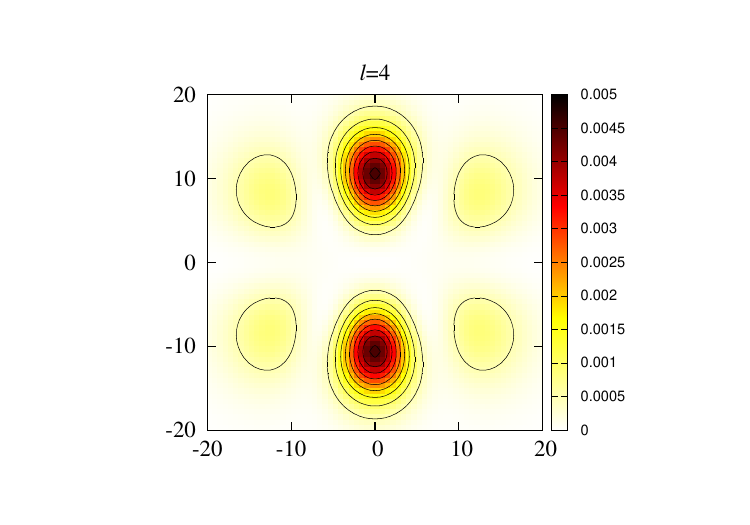,width=10cm}}
		\put(-1,6){\epsfig{file=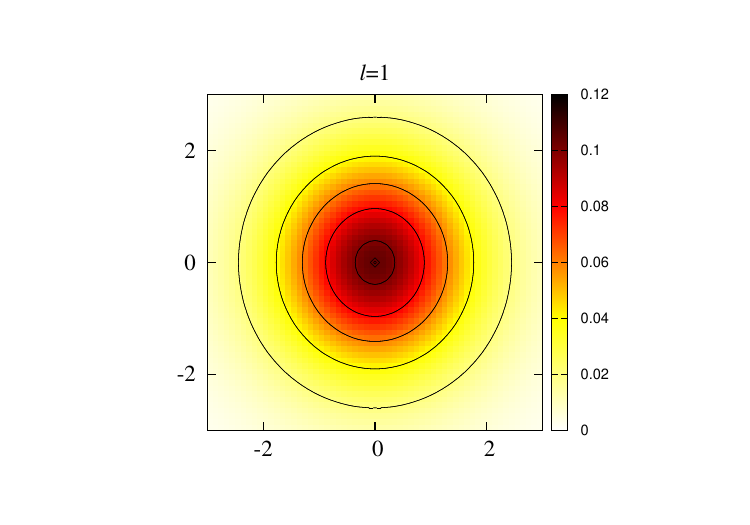,width=10cm}}
		\put(7,6){\epsfig{file=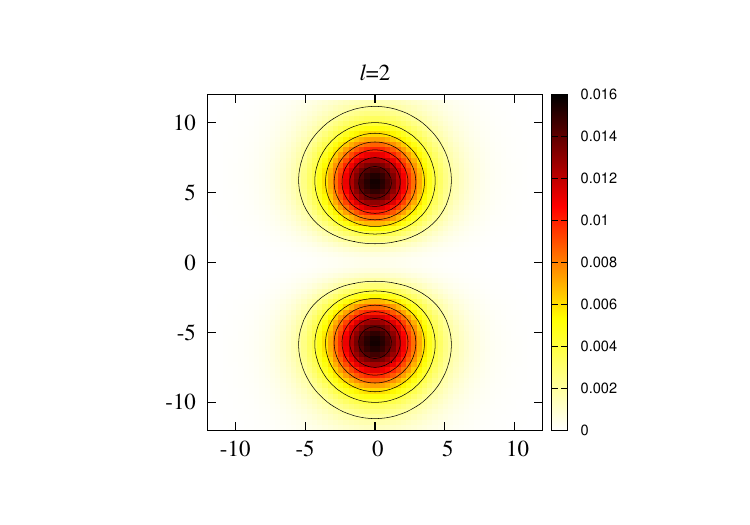,width=10cm}}
	\end{picture}
	\caption{
		Same as Figure \ref{En-magnetic} for
		typical electric solutions
		with $\ell=1,2,3,4$.
	}
    \label{En-electric}
\end{figure}

 The typical surfaces of constant energy density possess also an
 interesting morphology, as one can see in Figures
 \ref{En-magnetic} and \ref{En-electric}.
 In the magnetic case,
 they describe  $\ell$-coaxial tori, located symmetric $w.r.t$
 on the $z$-axis; 
 thus for an odd $\ell$, one of these tori is located in the equatorial plane - Figure \ref{En-magnetic}.

%
The situation with the electric case  is more involved - Figure~\ref{En-electric}.
 For $\ell=1$ the solutions are $prolate$ stars, with a single component \cite{Herdeiro:2023wqf}.
 For $\ell=2$ the energy is located in two centers, located symmetric on the $z-$axis,
 while for higher $\ell$ one finds in addition $\ell-2$ tori. 

 \begin{figure}[ht!]
 	\begin{center}
 		\includegraphics[height=.34\textwidth, angle =0 ]{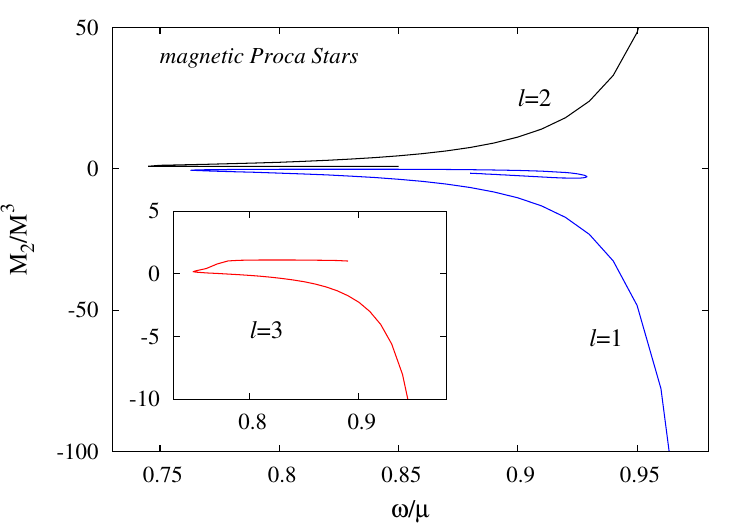}
 		\includegraphics[height=.34\textwidth, angle =0 ]{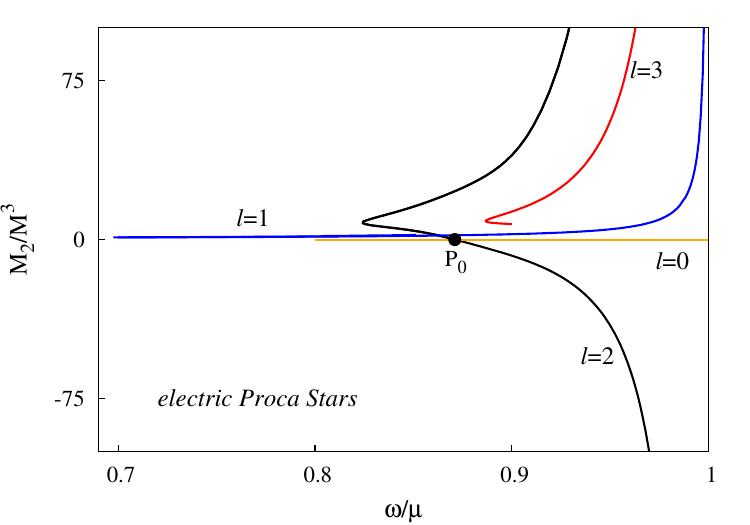}
 	\end{center}
 	\caption{  
 		{\small
 			The (reduced) quadrupole mass-moment is shown 
 			as a function of frequency.
 		}
 	}
 	\label{quadrupole}
 \end{figure}
 %
 %
 %

 \medskip
 
 The reduced quadrupole of the considered  families of solutions is shown 
  as a function of frequency
 in Figure \ref{quadrupole}.
For electric $\ell=2$ Proca stars, the quadrupole  
 changes sign at $P_0$, where it crosses the $M_2=0$ line for $\ell=0$ 
 spherical solutions, with the above mentioned bifurcation.
Although  the quadrupole moment also changes sign for $\ell=3$
 magnetic Proca stars, in that case 
 this feature could not be associated to any distinctive feature.

\begin{figure}[h!]
 	\begin{center}
    \includegraphics[height=.34\textwidth, angle =0 ]{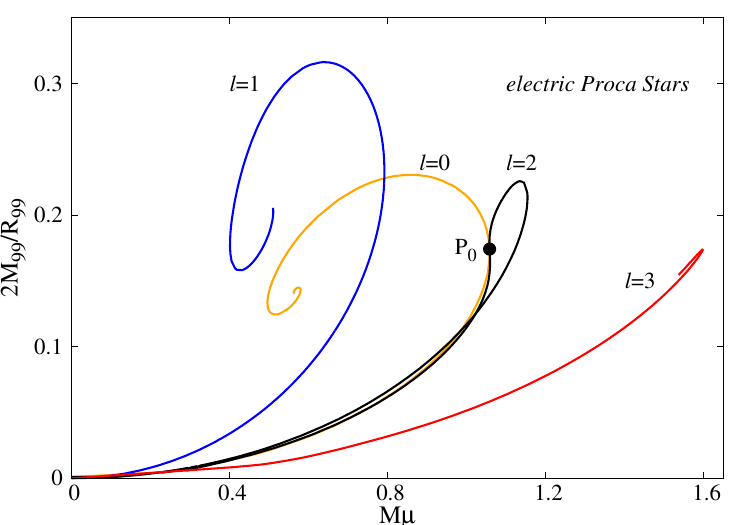}
 		\includegraphics[height=.34\textwidth, angle =0 ]{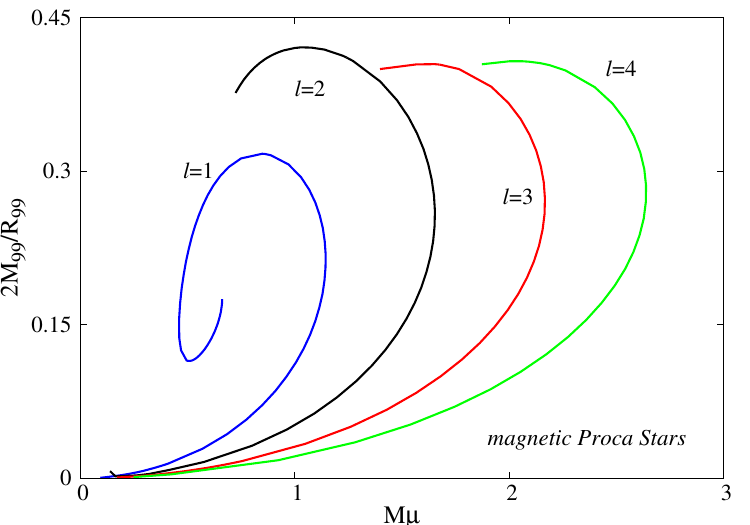}
 	\end{center}
 	\caption{  
 		{\small
 			The compactness is shown as a function of mass.
 		}
 	}
 	\label{comp}
 \end{figure}

  Finally, the
  compactness of the solutions is
  shown\footnote{$R_{99}$ is the perimetral radius containing 99\%
of the mass, $M_{99}$.} in Figure \ref{comp}.
  As one can see, 
  in the electric case the compactness attains its maximum for prolate stars, whereas in the magnetic case, the maximum compactness appears to saturate when increasing $\ell$.
 
\section{ Hybrid Proca stars }
\label{sec:hybrid}

\subsection{ General remarks }

Typically, the static  solitons allow for spinning generalizations,
which in many cases are disconnected from the static sector, without a slowly rotating limit.

The spinning PSs can be studied for   
a generalization of the line-element (\ref{metric})
with an extra-function associated with rotation, 
\begin{eqnarray}
	\label{metricR}
	ds^2=-e^{2F_0(r,\theta)}   dt^2+  e^{2F_1(r,\theta)} (dr^2+ r^2  d\theta^2) 
	+e^{2F_2(r,\theta)}r^2 \sin^2\theta 
	(d\varphi-\frac{W(r,\theta)}{r} dt)^2,
\end{eqnarray} 
the total angular 
momentum being read from
the far field asymptotics, with
$g_{\varphi t}=  -\frac{2J}{r}\sin^2\theta+\dots $.
The boundary conditions for the metric functions $F_i$ ($i=0,1,2$) are the same as in the 
static case, while for the metric function one imposes
with
\begin{eqnarray}
	\label{BCr1}
	W\big|_{r=0}=0,~ 
	W\big|_{r=\infty}=0,~
	\partial_\theta W\big|_{\theta=0,\pi}=0.
\end{eqnarray}
We assume that 
the geometry possesses the usual north-south (reflection) symmetry,
and thus the following boundary conditions are satisfied
in the equatorial plane:
   \begin{eqnarray}
   	\label{BCr3}
   	\partial_\theta F_i\big|_{\theta= \pi/2}=
    \partial_\theta W \big|_{\theta= \pi/2}=0.
   \end{eqnarray}

The Proca Ansatz
employed in all previous studies
on spinning Einstein-Proca solutions
contains 
a $\varphi$-dependence of the field's phase, with 
\begin{eqnarray}
	\label{Paxial}
\mathcal{A}=\left(
	\frac{H_1}{r}dr+H_2d\theta+i H_3 \sin \theta d\varphi + iVdt   
	\right)
	e^{i(m\varphi-\omega t)} \, , \ \ \ \ \  
\end{eqnarray}
where $m \in \mathbb{Z}^+$.
The presence of both electric and magnetic parts in the axially symmetric
Ansatz (\ref{Paxial})
results in a non-zero  angular momentum density stored in the Proca field, with 
\begin{eqnarray}
	\label{s1}
	T_\varphi^t=\frac{1}{2}
	\left [
	{\mathcal{F}}_{\alpha \varphi} \bar{\mathcal{F}}^{\alpha t}+\bar{\mathcal{F}}_{\alpha \varphi}  {\mathcal{F}}^{\alpha t}
	+\mu^2 (
	\mathcal{A}_{\varphi} \bar{\mathcal{A}}^{t}
	+\bar{\mathcal{A}}_{\varphi}\mathcal{A}^{t}
	)
	\right ]  \ ,
\end{eqnarray}
which can be re-expressed as
\begin{eqnarray}
	\label{s3}
	T_\varphi^t  
	&=&\frac{1}{2}
	\left\{
	\frac{1}{\sqrt{-g}}\partial_\alpha \left[ ({\mathcal{A}}_\varphi \bar{ {\mathcal{F}}}^{\alpha t}+\bar{\mathcal{A}}_\varphi { {\mathcal{F}}}^{\alpha t}  )\sqrt{-g} \right]
	-i m 
	(
	\mathcal{A}_{\alpha} \bar{ {\mathcal{F}}}^{\alpha t}
	- 
	\bar{\mathcal{A}}_{\alpha}{ {\mathcal{F}}}^{\alpha t}
	)
	\right.
	\nonumber 
	\\
    &&
	\left. - \frac{1}{\sqrt{-g}}\left[{\mathcal{A}}_\varphi\partial_\alpha( \bar{ {\mathcal{F}}}^{\alpha t}\sqrt{-g})
	+\bar{\mathcal{A}}_\varphi \partial_\alpha({ {\mathcal{F}}}^{\alpha t} \sqrt{-g}) \right] 
	+\mu^2 (
	\mathcal{A}_{\varphi} \bar{\mathcal{A}}^{t}
	+\bar{\mathcal{A}}_{\varphi}\mathcal{A}^{t}
	)
	\right\} \ .
\end{eqnarray}

However, the Proca equations imply that
the second line above is identically zero,
and thus the  angular momentum density
can be written as
\begin{eqnarray}
	\label{s4}
	T_\varphi^t  =m j^t 
	+ \nabla_\alpha P^\alpha \ ,
\end{eqnarray}
with $ j^t $ the Noether charge density and
\begin{eqnarray}
	\label{s5}
	P^\alpha= {\mathcal{A}}_\varphi \bar{ {\mathcal{F}}}^{\alpha t}+\bar{\mathcal{A}}_\varphi { {\mathcal{F}}}^{\alpha t}   \ .
\end{eqnarray}
Then the following relation 
follows 
\cite{Herdeiro:2016tmi}
\begin{eqnarray}
	\label{JQ}
	J= \int T_\varphi^t  \sqrt{-g}d^3x=m Q ,
\end{eqnarray}
the contribution of the   boundary terms
(in terms of $P^\alpha$) vanishing for  regular configurations\footnote{The decomposition (\ref{s4})
of the angular momentum density
in two different 
components is a vector field feature.  We recall that 
$T_\varphi^t  =m j^t$
for a scalar field \cite{Schunck:1996he}.
}.

\subsection{ Local  rotation with vanishing global rotation}

Even for $m=0$, which by virtue of~\eqref{JQ} means $J=0$ and thus no global rotation, 
 turning on $both$ the  
electric and the magnetic potentials 
 yields  a non-zero
angular momentum density, since 
the functions
$ P^\alpha$ as given by (\ref{s5})
are  locally non-zero\footnote{This can easily be verified for the case of the flat
	space (singular) solution in Section \ref{section1}.}.
This suggests 
the existence of PSs which spin locally ($T_\varphi^t \neq 0$)
but not globally\footnote{This exotic feature is known to
occur also for spinning dyons in  Yang-Mills-Higgs model 
(with the Higgs field in the adjoint representation)
\cite{VanderBij:2001nm,Kleihaus:2005fs}, while it is absent for spinning solitons in the electroweak sector of the Standard Model 
\cite{Radu:2008ta,Kleihaus:2008cv}.
}.

\begin{figure}[h!]
\begin{center}
 \includegraphics[height=.34\textwidth, angle =0 ]{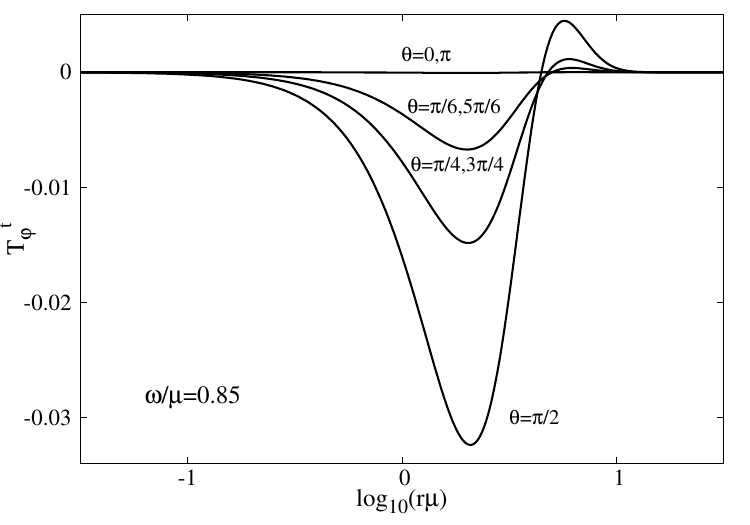}
 \includegraphics[height=.34\textwidth, angle =0 ]{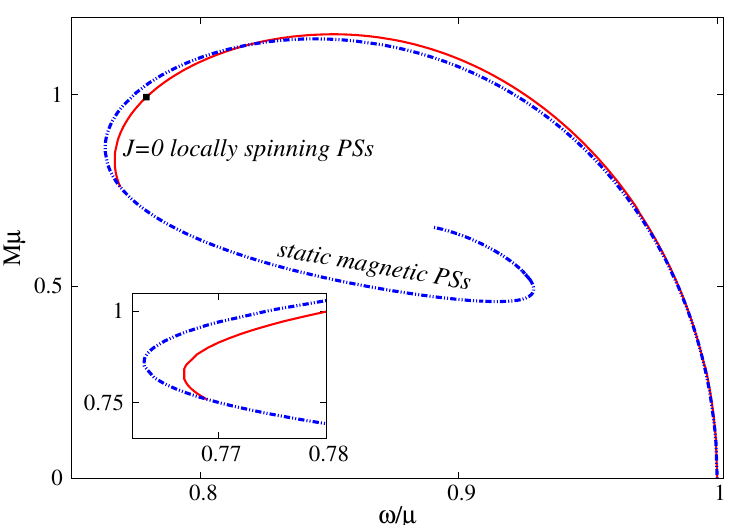}
\end{center}
\caption{ 
{\small 
{\it Left:}   
 The angular momentum density
$T_\varphi^t$
is shown as a function of radial coordinate
and several angles for a  
typical $J=0$  
 hybrid \{electric $\ell = 0 $, magnetic $\ell = 1$\}-spinning 
Proca star.
For any $\theta>0$, $T_\varphi^t$ becomes positive for large enough $r$,
such that its volume integral
 vanishes, $J=0$. 
{\it Right:} 
The frequency-mass diagram is shown for the $J=0$ family of  
spinning Proca stars (PSs)
consisting in a non-linear superposition of
an electric monopole  and a magnetic dipole.
The angular momentum density vanishes for a critical frequency, smoothly joining 
the family of static,  magnetic, $\ell=1$ configurations. 
}
}
\label{wM-rot}
\end{figure}

To address this issue,
let us consider a {\it mixed} Proca ansatz, with
$m=0$ (no $\varphi$-dependence) and
\begin{eqnarray}
\label{ansatzGen}
\mathcal{A}=\mathcal{A}_m+\mathcal{A}_e,
\end{eqnarray}
where $\mathcal{A}_m$ and $\mathcal{A}_e$
are given by (\ref{ansatzM}) and (\ref{ansatzE}), respectively,
and look for solitonic solutions of the Einstein-Proca equations.
We have found numerical evidence for the existence of
a set of such solutions
which can be thought as 
a non-linear superposition of the
$electric$ $\ell=0$ 
and
$magnetic$ $\ell=1$
configurations discussed in Section \ref{solutions}.
The boundary conditions for the Proca potential functions are, in this case,
\begin{eqnarray}
	\label{BCn2}
	H_i|_{r=0}=\partial_r V|_{r=0}=0,~H_i|_{r=\infty}=V|_{r=\infty}=0,~
	\partial_\theta H_1|_{\theta=0,\pi}= H_2\big|_{\theta=0,\pi}=  H_3\big|_{\theta=0,\pi}
	=\partial_\theta V|_{\theta=0,\pi}=0~,
\end{eqnarray} 
while in the equatorial plane we impose
\begin{eqnarray}
	\label{BCn3}
	\partial_\theta  H_1|_{\theta=\pi/2}=
	H_2\big|_{\theta=\pi/2}= \partial_\theta  H_3\big|_{\theta=\pi/2}=     \partial_\theta  V|_{\theta=\pi/2}=0\ .
\end{eqnarray}
All these solutions still possess a $\mathbb{Z}_2$-symmetry. 

In Figure  \ref{wM-rot} 
(left panel) we show the angular momentum density for several  different values of $\theta$.
One notices that, while $T_{\varphi}^t<0$ in a small region close to the origin,
it changes sign for large enough $r$, such that the volume integral of $T_{\varphi}^t$
 vanishes.

In a ADM mass $vs.$  frequency diagram,
Figure \ref{wM-rot}  
(right panel), 
 the spinning solutions form a curve which starts again at 
$\omega=\mu$ where $M=0$.  
 As one moves up
along the curve, the ADM mass increases and $\omega$ decreases, until a maximal mass is attained.
There is also a minimal frequency, with the emerging of a new branch of solutions,
which, however, stops to exist for a frequency 
 $\omega_{critical} \simeq 0.76893$.
There the metric function $W$
and the Proca potentials 
$H_1,H_2,V$
vanish,
and the set of spinning solutions 
smoothly joins the family of static magnetic solutions with
$\ell=1$.
A similar picture is found for the Noether charge,
with $M < \mu Q$
for the region between
the maximal frequency
and a configuration
marked with a black dot in  
Figure  \ref{wM-rot} 
(right panel). 

\begin{figure}[h!]
\begin{center}
 \includegraphics[height=.34\textwidth, angle =0 ]{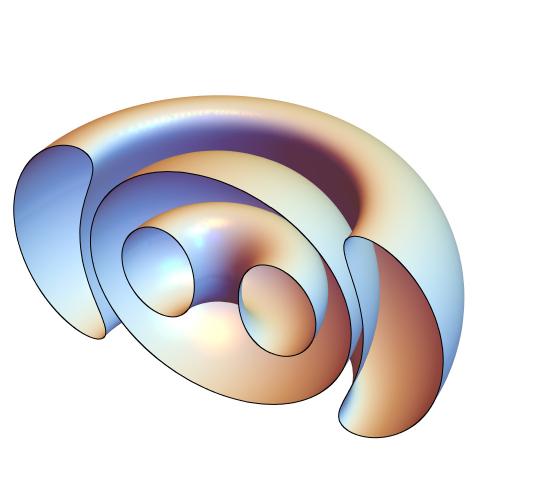}
 \includegraphics[height=.34\textwidth, angle =0 ]{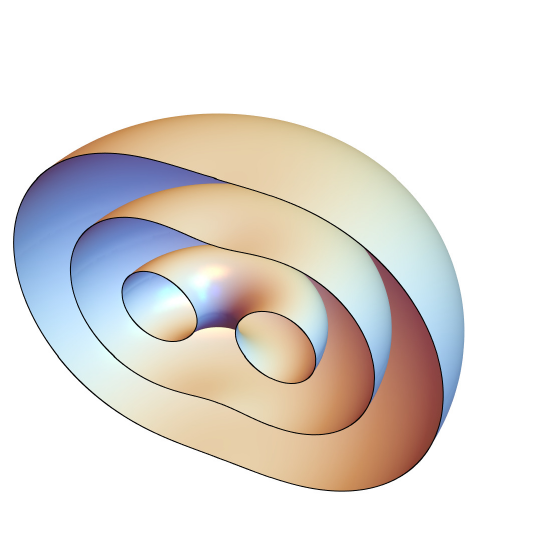}
\end{center}
\caption{ 
{\small 
Surfaces of constant angular momentum 
density
(left panel)
and energy density
(right panel)
for a  
$J=0$  
 hybrid \{electric $\ell = 0 $, magnetic $\ell = 1$\}-spinning 
Proca star with $\omega/\mu=0.8$.
The considered values are
$T_\varphi^t=0.004$ (outside torus),$-0.03$ (inside torus) and $0$ (the squashed sphere). 
For energy densities we consider are
$-T_t^t=0.05,0.02, 0.005$ (from inside to outside).
}
}
\label{wM-rot3}
\end{figure}

In Fig.~\ref{wM-rot3} we illustrate the surfaces of constant angular momentum density
(left panel)
and energy density
(right panel)
for a typical $J=0$ 
spinning Proca star belonging to this family.

Similar solutions exist for other
choices of the  electric and magnetic 
$\ell$-fundamental modes
and considering again their non-linear superposition.
In particular, the configurations with 
both values of $\ell$
odd or even result in spinning Proca stars {\it without}
a $\mathbb{Z}_2$-symmetry.\footnote{This is a feature of some known black hole solutions in specific non-vacuum models, see $e.g.$ \cite{Cunha:2018uzc,Cunha:2024gke} and references therein. 
However, we are not aware of any axially symmetric solitonic solution  without a $\mathbb{Z}_2$-symmetry.
} 
This can be seen already
when considering the energy-momentum tensor and the Noether charge density of the fundamental modes.

As an illustration of non-$\mathbb{Z}_2$-symmetric solutions, we briefly discuss here solutions with both electric and magnetic $\ell=1$ modes. 
They are constructed by employing the same approach as before, with the Ansatz 
(\ref{metricR}), (\ref{ansatzGen}).
However, this time one considers in the numerics
the full interval
$0\leq \theta \leq \pi$, with the
boundary conditions
\begin{eqnarray}
 	\partial_\theta H_1|_{\theta=0,\pi}= H_2\big|_{\theta=0,\pi}=  H_3\big|_{\theta=0,\pi}
	=\partial_\theta V|_{\theta=0,\pi}=0~.   
\end{eqnarray}
It is convenient to employ a new compactified radial coordinate $x$,
with
$r= c x^2/(1-x^2)$, where $c=5$ for most of the solutions here.
The imposed boundary conditions at the ends of the $x$-interval are 
\begin{eqnarray}
	\partial_x F_i\big|_{x=0} =W\big|_{x=0}=  0,~~ F_i\big|_{x=1}=W\big|_{x=1}  =0, ~~ 
	H_i|_{x=0}=\partial_x V|_{x=0}=0,~H_i|_{x=1}=V|_{x=1}=0.
\end{eqnarray}

\begin{figure}[h!]
\begin{center}
 \includegraphics[height=.34\textwidth, angle =0 ]{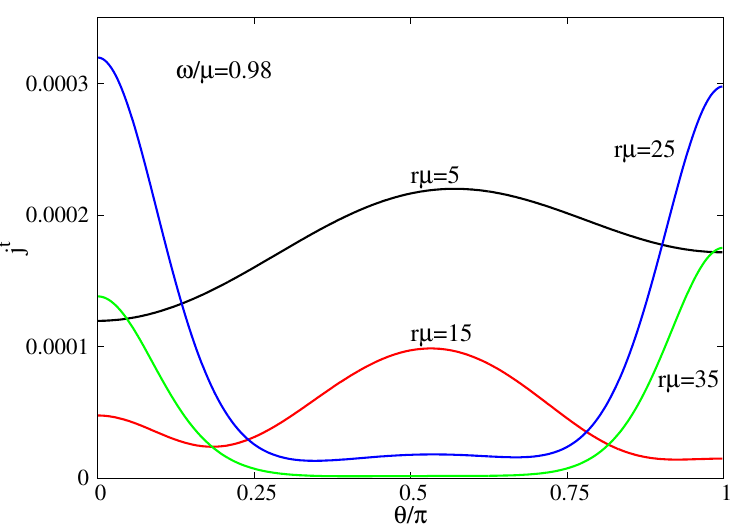}
 \includegraphics[height=.34\textwidth, angle =0 ]{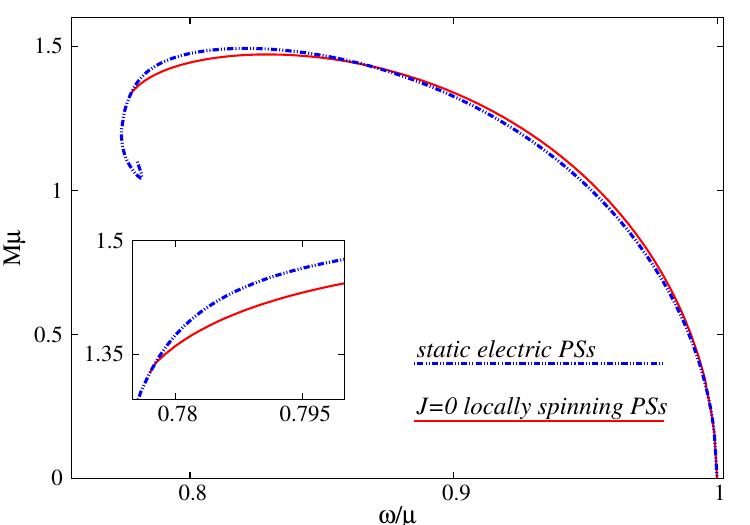}
\end{center}
\caption{ 
{\small 
{\it Left:}   
 The Noether charge density
$j^t$
is shown as a function of angular variable $\theta$
and several values of the radial coordinate for a typical $J=0$  
 hybrid \{electric $\ell = 1 $, magnetic $\ell = 1$\}-spinning Proca star.
 One notices the absence of a reflection symmetry with respect to 
 $\theta=\pi/2$.
{\it Right:} 
The frequency-mass diagram is shown for the $J=0$ family of  
spinning Proca stars (PSs)
consisting in a non-linear superposition of
an electric dipole  and a magnetic dipole.
The angular momentum density vanishes for a critical frequency, smoothly joining 
the family of static,  electric, $\ell=1$ configurations possessing one node for all Proca
potentials. 
}
}
\label{wM-rot2}
\end{figure}

The ADM mass $vs.$  frequency diagram of these solutions is similar to that displayed in  
Fig.~\ref{wM-rot}  
for the  hybrid \{electric $\ell = 0 $, magnetic $\ell = 1$\} configurations, being shown in Fig.~\ref{wM-rot2}.
However,
in contrast to that case, 
 all hybrid  \{electric $\ell = 1 $, magnetic $\ell = 1$\} solutions
 we have been able to construct
 possess a node for the  potentials $H_1$, $H_2$ and $V$.
 Another difference is that 
 they bifurcate for a critical frequency
 $\omega_{critical} \simeq 0.7776 \mu$
 from 
 the family of static electric solutions with 
 $\ell=1$
 and one node for all Proca potentials\footnote{The fundamental solutions electric $\ell = 1 $ Proca stars
 reported in the literature (and also in the previous Section) are nodeless.
 We also mention that we have found  $\ell = 1 $ electric solutions
 with nodes for some of the potentials, only.}.

 %

The mechanism of ensuring $J=0$ is similar to the previous set of solutions,
with $T_\varphi^t $ changing sign, so that its volume integral  
vanishes.
Their distinctive feature - the violation of $\mathbb{Z}_2$-symmetry - is barely noticeable  for most of the solutions -
 see Fig.~\ref{wM-rot2} (left panel),
 where the Noether charge density is shown for a typical configuration.

Finally,
we remark that all $J=0$ spinning Proca Stars 
constructed so far
do not possess an ergoregion.

\section{Dynamical evolutions and stability}
\label{sec:evolutions}

It is important to assess the dynamical stability of the new magnetic and hybrid solutions reported in the previous sections. We now describe the corresponding analysis.

\subsection{Setup}

We have performed fully numerical-relativity simulations of the Einstein-(complex) Proca system to address the dynamical stability of the new multipolar Proca stars.  Adopting the standard $3+1$ decomposition (see, e.g.,~\cite{Zilhao:2015tya,Sanchis-Gual:2017bhw,Sanchis-Gual:2019ljs} for further details), the Proca field can be expressed in terms of the following $3+1$ variables:
\begin{eqnarray}
\mathcal{A}_{\mu} &=& \mathcal{X}_{\mu} + n_{\mu}\mathcal{X}_{\phi}, \\
\mathcal{X}_{i}   &=& \gamma^{\mu}{}_{i}\,\mathcal{A}_{\mu}, \\
\mathcal{X}_{\phi} &=& - n^{\mu}\mathcal{A}_{\mu},
\end{eqnarray}
where $n^{\mu}$ denotes the timelike unit normal vector to the spatial hypersurfaces, and $\gamma^{\mu}{}_{\nu} = \delta^{\mu}{}_{\nu} + n^{\mu} n_{\nu}$ is the projection operator onto these hypersurfaces. The quantity $\mathcal{X}_{i}$ corresponds to the spatial vector potential, while $\mathcal{X}_{\phi}$ represents the scalar potential. The time evolution of the Einstein-Proca system is described by the following set of equations
\begin{eqnarray}
\partial_{t} \gamma_{ij} & = & - 2 \alpha K_{ij} + \mathcal{L}_{\beta} \gamma_{ij}, \label{eq:dtgamma}\\
\partial_{t} \mathcal{X}_{i}      & = & - \alpha \left( E_{i} + D_{i} \mathcal{X}_{\phi} \right) - \mathcal{X}_{\phi} D_{i}\alpha + \mathcal{L}_{\beta} \mathcal{X}_{i}, \label{eq:dtAi}\\
\partial_{t} E^{i}       & = &
        \alpha \left( K E^{i} + D^{i} Z + \mu^2 \mathcal{X}^{i}
                + \epsilon^{ijk} D_{j} B_{k} \right) \nonumber \\
        & & - \epsilon^{ijk} B_{j} D_{k}\alpha
        + \mathcal{L}_{\beta} E^{i}, \label{eq:dtEi}\\
\partial_{t} K_{ij}      & = & - D_{i} D_{j} \alpha
        + \alpha \left( R_{ij} - 2 K_{ik} K^{k}{}_{j} + K K_{ij} \right)
\nonumber \\
& & + 2 \alpha \biggl( E_{i} E_{j} - \frac{1}{2} \gamma_{ij} E^{k} E_{k}
        + B_{i} B_{j} - \frac{1}{2} \gamma_{ij} B^{k} B_{k}
        - \mu^{2} \mathcal{X}_{i} \mathcal{X}_{j} \biggr)
        + \mathcal{L}_{\beta} K_{ij}, \label{eq:dtK}\\
\partial_{t} \mathcal{X}_{\phi}  & = & - \mathcal{X}^{i} D_{i} \alpha
        + \alpha \left( K \mathcal{X}_{\phi} - D_{i} \mathcal{X}^{i} - Z \right)
        + \mathcal{L}_{\beta} \mathcal{X}_{\phi}, \label{eq:dtphi}\\
\partial_{t} Z          & = & \alpha \left( D_{i} E^{i} + \mu^{2} \mathcal{X}_{\phi} - \kappa Z \right)
        + \mathcal{L}_{\beta} Z\label{eq:dtZ}.
\end{eqnarray}

Here, $\alpha$ denotes the lapse function, $\beta^{i}$ the shift vector, and $\gamma_{ij}$ the induced spatial metric. The extrinsic curvature is represented by $K_{ij}$, with trace $K = K^{i}{}_{i}$, while $D_{i}$ indicates the covariant derivative compatible with $\gamma_{ij}$. The operator $\mathcal{L}_{\beta}$ corresponds to the Lie derivative along the shift vector. The parameter $\kappa$ is introduced as a damping coefficient to enhance numerical stability.

In addition, one defines the three-dimensional ``electric'' and ``magnetic'' fields, $E^{i}$ and $B^{i}$, in close analogy with Maxwell theory:
\begin{equation}
E_{i} = \gamma^{\mu}{}_{i}\,\mathcal{F}_{\mu\nu} n^{\nu}, 
\qquad 
B_{i} = \gamma^{\mu}{}_{i}\,{}^{\star}\mathcal{F}_{\mu\nu} n^{\nu}
      = \epsilon^{ijk} D_{j} \mathcal{X}_{k},
\end{equation}
where both fields are purely spatial, satisfying $E_{\mu} n^{\mu} = B_{\mu} n^{\mu} = 0$, and $\epsilon^{ijk}$ denotes the Levi-Civita tensor associated with the spatial hypersurfaces. The Hamiltonian constraint
and the momentum constraint  are given by:
\begin{eqnarray}
\label{eq:Hamiltonian}
\mathcal{H}  &=& R - K_{ij} K^{ij} + K^2 
        - 2 \bigl( E^{i} E_{i} + B^{i} B_{i}+  \mu^{2} \bigl( \mathcal{X}_{\phi}^{2} 
        + \mathcal{X}^{i} \mathcal{X}_{i} \bigl)
            \bigl)= 0\,,\\
\label{eq:momentumConstraint}
\mathcal{M}_{i}  &=& D^{j} K_{ij} - D_{i} K 
        - 2 \bigl( \epsilon_{ijk} E^{j} B^{k} +\mu^{2} \mathcal{X}_{\phi} \mathcal{X}_{i}
            \bigl)
       = 0
\,.
\end{eqnarray}
 
The computational domain is specified by
\begin{equation}
x_{\min} = y_{\min} = z_{\min}= -99.0,
\end{equation}
\begin{equation}
x_{\max} = y_{\max} = z_{\max} = 99.0.
\end{equation}
No symmetry conditions are imposed. The grid spacing on the coarsest level is
\begin{equation}
\Delta x = \Delta y = \Delta z = 1.1.
\end{equation}
A fixed mesh refinement strategy with three levels is employed. The grid hierarchy is given by
\begin{equation}
\{(99, 49.5, 24.75)/\mu,\; (1.1, 0.55, 0.275)/\mu\},
\end{equation}
for the magnetic Proca star case, while for the hybrid case we have used
\begin{equation}
\{(99, 49.5, 24.75)/\mu,\; (0.99, 0.495, 0.2475)/\mu\},
\end{equation}
where the first set represents the spatial extent of each refinement level and the second set indicates the corresponding resolution.

The numerical simulations are performed using the publicly available \texttt{Einstein Toolkit}~\cite{EinsteinToolkit:2024_11,loffler2012einstein}, which is built upon the \texttt{Cactus} computational framework~\cite{Goodale:2002a} and supports adaptive mesh refinement. Time integration of the evolution equations is carried out via the method of lines, employing a fourth-order Runge--Kutta scheme. The geometric sector, corresponding to the left-hand side of the Einstein equations, is evolved using the \texttt{McLachlan} code~\cite{brown2009turduckening,reisswig2011gravitational}, which implements the $3+1$ Baumgarte--Shapiro--Shibata--Nakamura (BSSN) formulation. The Proca field evolution equations, Eqs.~(\ref{eq:dtgamma})--(\ref{eq:dtZ}), are solved with the infrastructure introduced in~\cite{Zilhao:2015tya,witek_2023_7791842}. This framework has been further extended to accommodate a complex Proca field~\cite{Sanchis-Gual:2018oui,Sanchis-Gual:2019ljs,Sanchis-Gual:2020mzb,Sanchis-Gual:2021edp, Sanchis-Gual:2022mkk,Herdeiro:2023wqf, Lazarte:2025wlw, Palloni:2025mhn}. Additional implementation details, code validation, and convergence analyses can be found in~\cite{Zilhao:2015tya,Sanchis-Gual:2019ljs,Sanchis-Gual:2022mkk,Lazarte:2025wlw,Palloni:2025mhn}. No explicit perturbations are added to the initial data. As a result, the  underlying instabilities are triggered by numerical truncation errors.

\subsection{Magnetic stars ($\ell=1$)}

We begin by evolving the $\ell=1$ magnetic Proca star configurations introduced in the previous sections. In order to explore different regimes, we consider four representative models given by
\begin{equation}
(\omega/\mu, M\mu) = \{(0.98, 0.570), (0.95, 0.849), (0.90, 1.072), (0.85, 1.144)\}.
\end{equation}
These configurations cover both the tentative stable (i.e. from maximal frequency to maximal mass) and unstable (the remaining existence curve) regions of the solution space.

The most massive model is located beyond the maximum mass supported by this family and belongs to the radially unstable branch. As a result, it undergoes gravitational collapse and forms a black hole. The remaining configurations do not collapse, but instead develop a  non-axisymmetric instability that drives their evolution. In particular, they evolve toward prolate Proca star configurations, in agreement with previous results for spherical Proca stars~\cite{Herdeiro:2023wqf,Lazarte:2025wlw}.

\begin{figure}[h!]
\begin{center}
 $xy$-plane\\
 \includegraphics[height=.4\textwidth, angle =0 ]{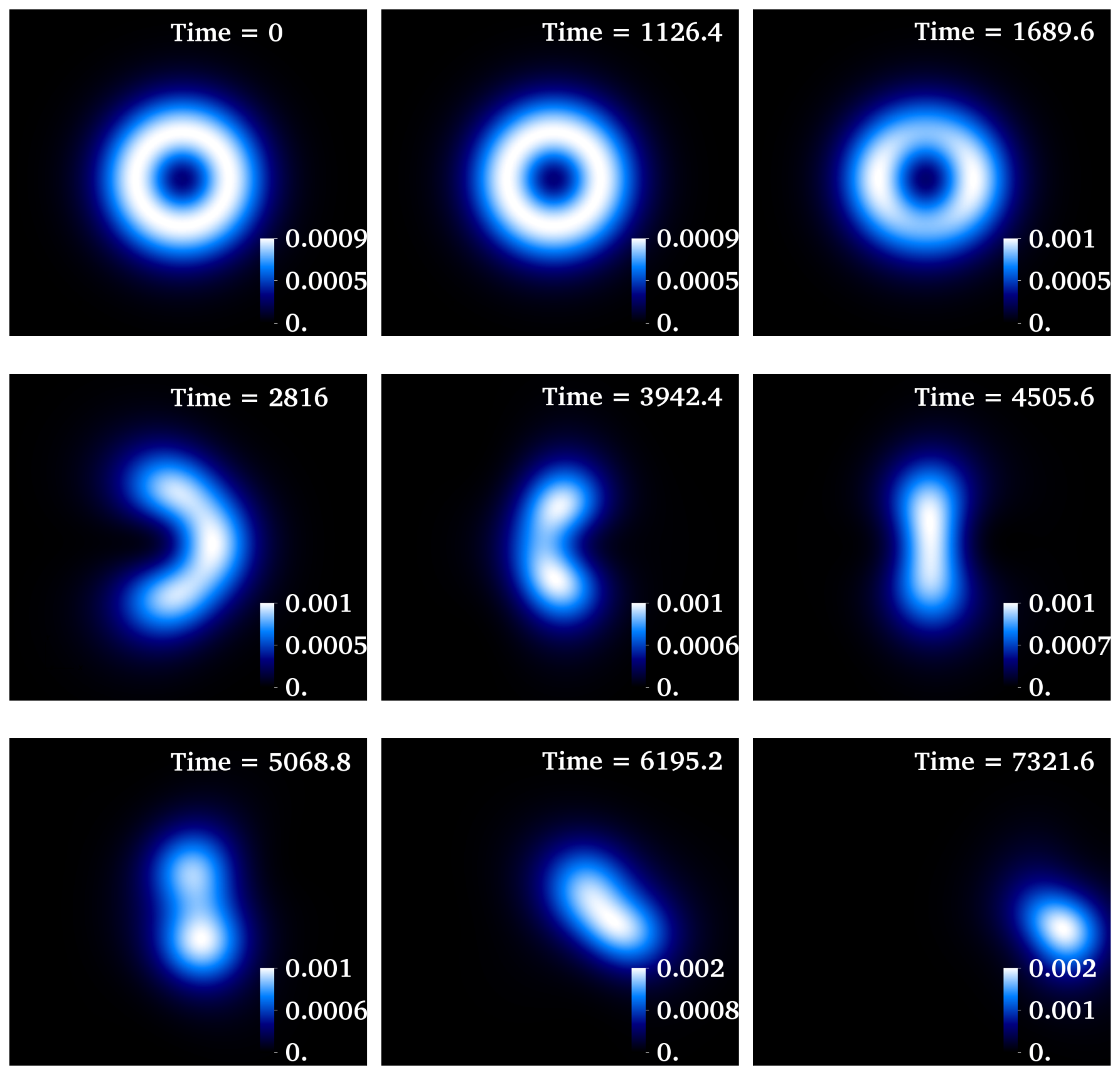}
 \includegraphics[height=.4\textwidth, angle =0 ]{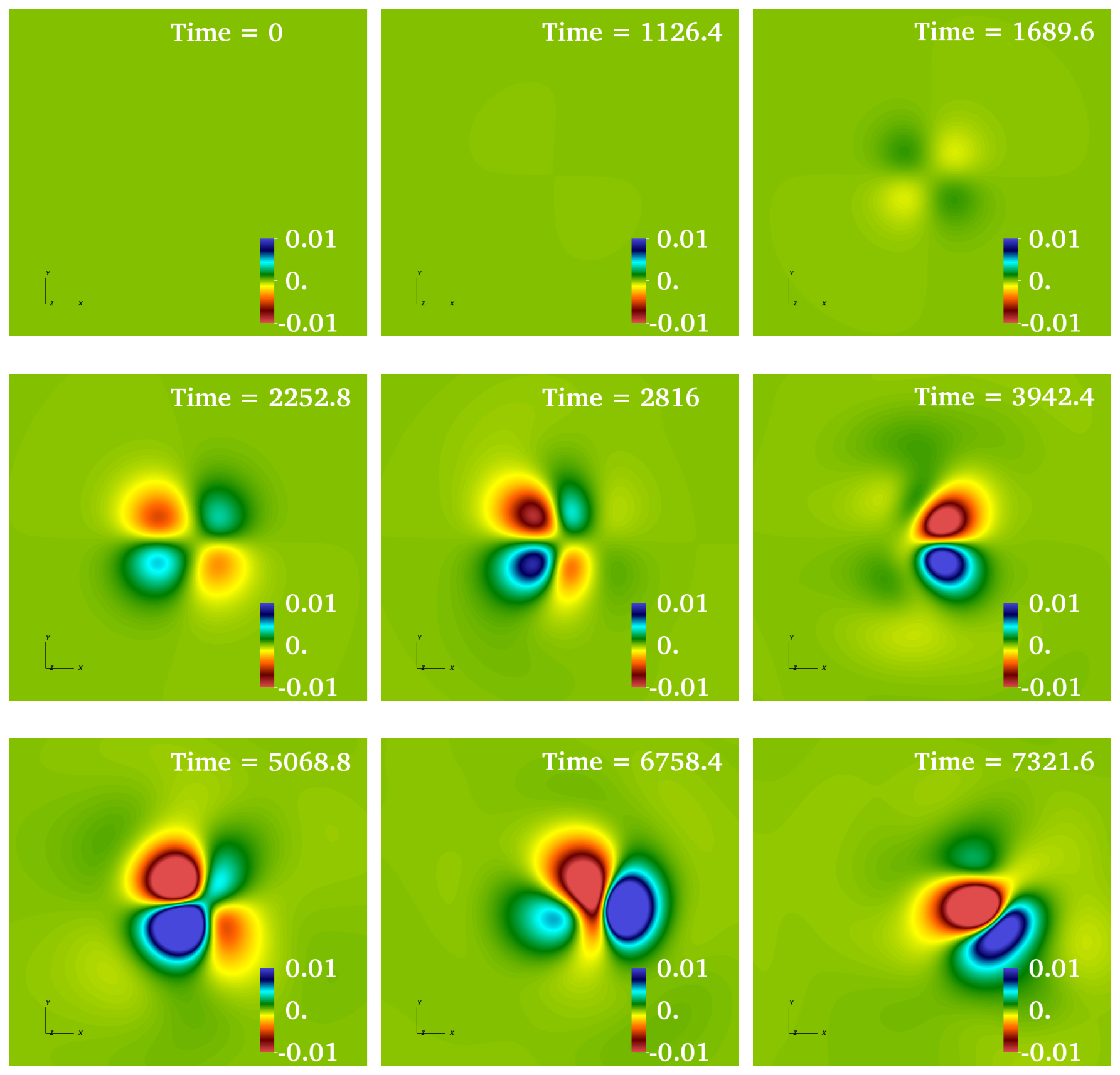}\\
$xz$-plane\\
 \includegraphics[height=.4\textwidth, angle =0 ]{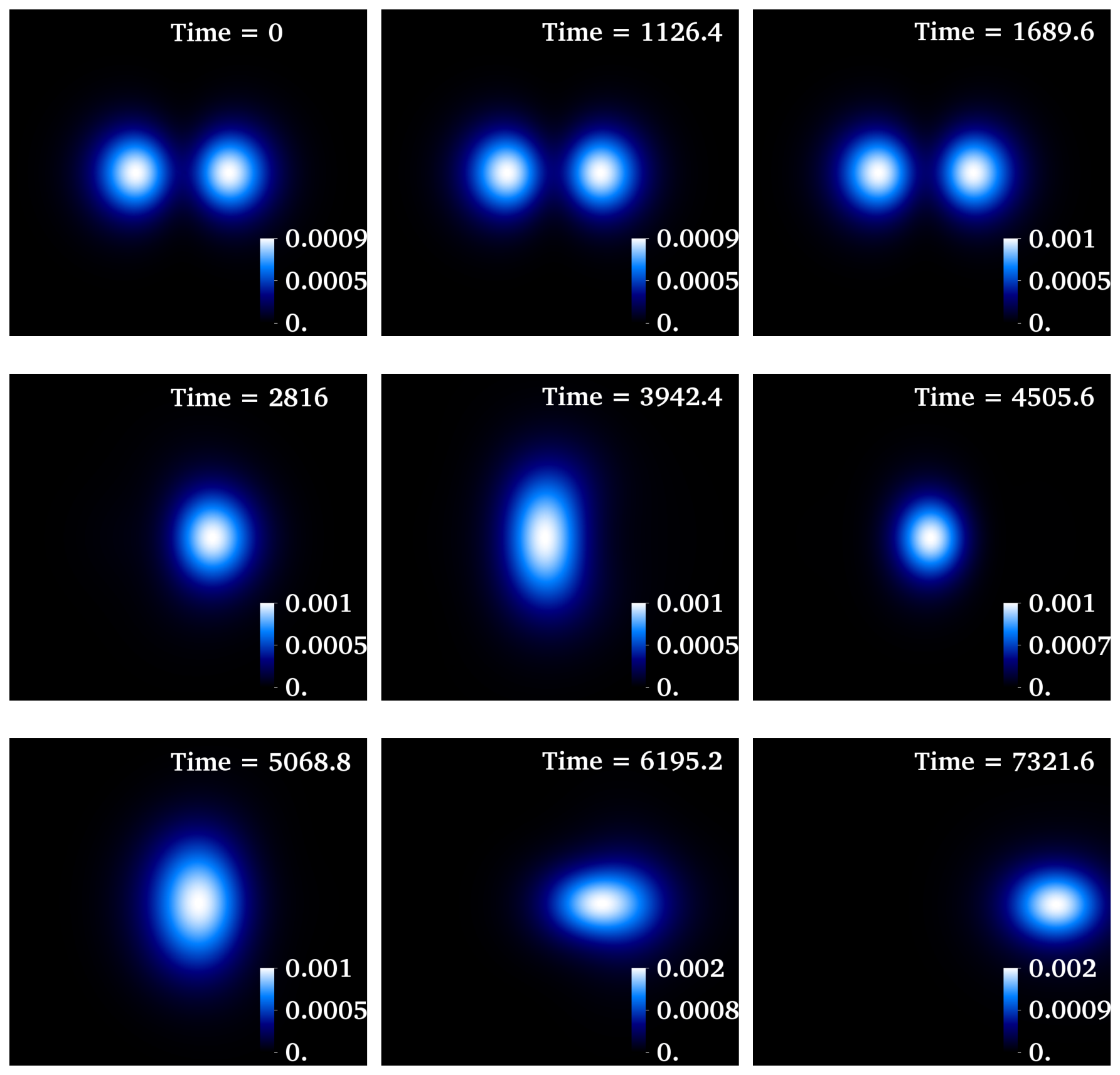}
 \includegraphics[height=.4\textwidth, angle =0 ]{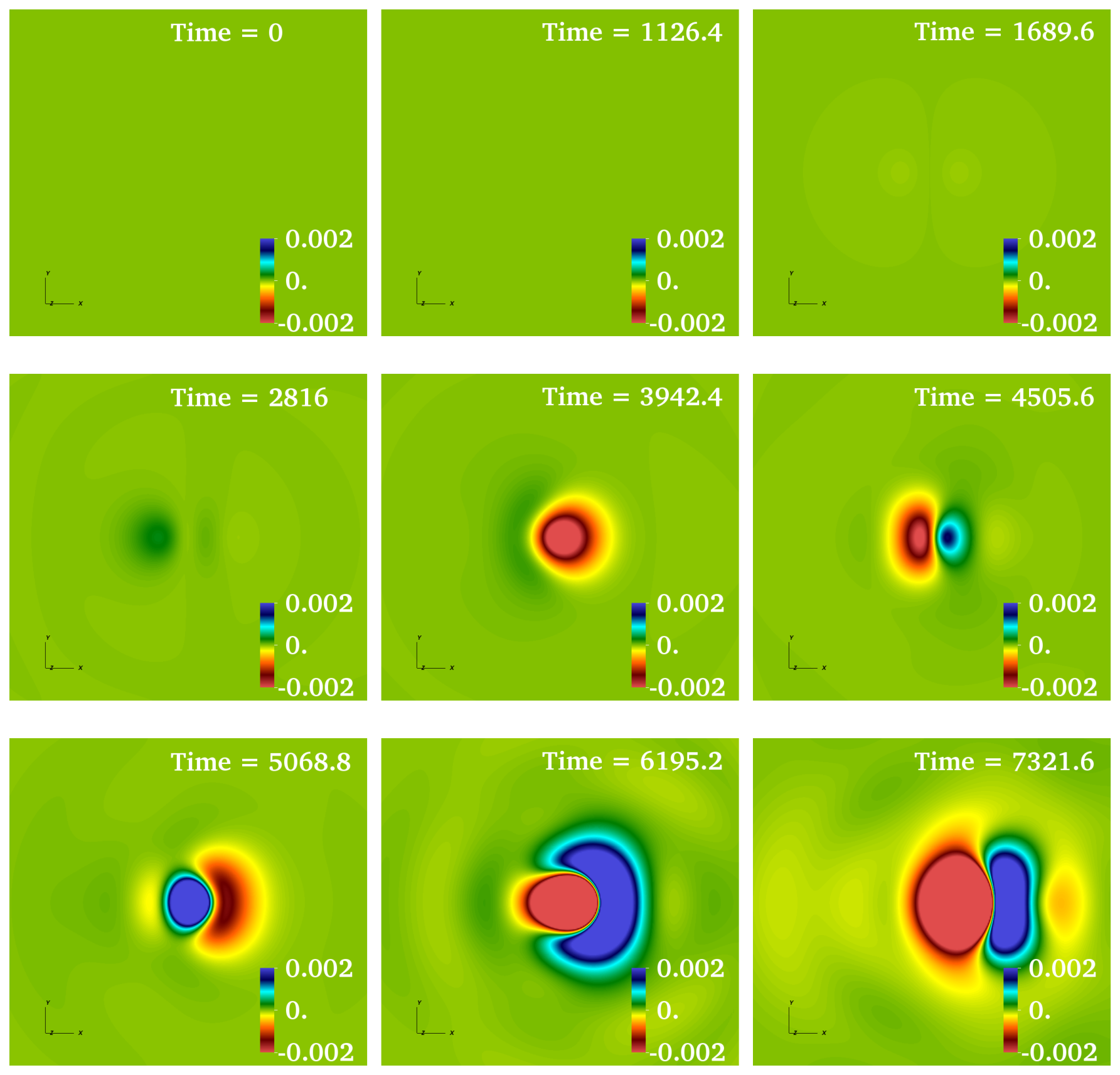}
\end{center}
\caption{ 
{\small 
{\it Left:}   
 Snapshots of the time evolution of the Proca Komar energy density in the $xy$ and $xz$ planes for the magnetic Proca star with $\omega/\mu = 0.90$. Time increases from left to right and from top to bottom.
\textit{Right:} Same as in the left panels, but for the real part of the scalar potential $\mathcal{X}_{\phi}$. 
}
}
\label{evol_l1}
\end{figure}

Further details of this evolution are shown in Fig.~\ref{evol_l1}, where we present the Komar energy density and the real part of the scalar potential $\mathcal{X}_{\phi}$ in the $xy$ and $xz$ planes. At early times, the stars preserve their initial toroidal structure. However, the system quickly develops an $m=2$ mode, followed by the growth of an $m=1$ mode around $t\mu \sim 1500$. This leads to the breakdown of the toroidal morphology and drives the system toward a configuration with the maximum of the energy density located at the center.

The behavior of the scalar potential provides additional information about this transition. Initially, and in agreement with the magnetic ansatz in Eq.~(\ref{ansatzM}), $\mathcal{X}_{\phi}$ vanishes. This feature is clearly visible at early times. As the evolution proceeds and the system departs from the initial configuration, $\mathcal{X}_{\phi}$ develops a dipolar ($m=1$) structure. This supports the interpretation of the final state as a prolate configuration.

\subsection{Hybrid stars}

We now turn to the hybrid case, where we superimpose electric and magnetic solutions to obtain rotating solutions. We will study two cases: $\ell=0$ electric + $\ell=1$ magnetic and $\ell=1$ electric + $\ell=1$ magnetic.

\subsubsection{$\ell=0$ electric + $\ell=1$ magnetic}

We have evolved three representative configurations characterized by
\begin{equation}
(\omega/\mu, M\mu) = \{(0.98, 0.573), (0.95, 0.859), (0.93, \mathrm{0.977})\}.
\end{equation}
These solutions possess non-vanishing local angular momentum density, indicating local rotation, while their total angular momentum vanishes (zero global rotation).

\begin{figure}[h!]
\begin{center}
 \includegraphics[height=.8\textwidth, angle =0 ]{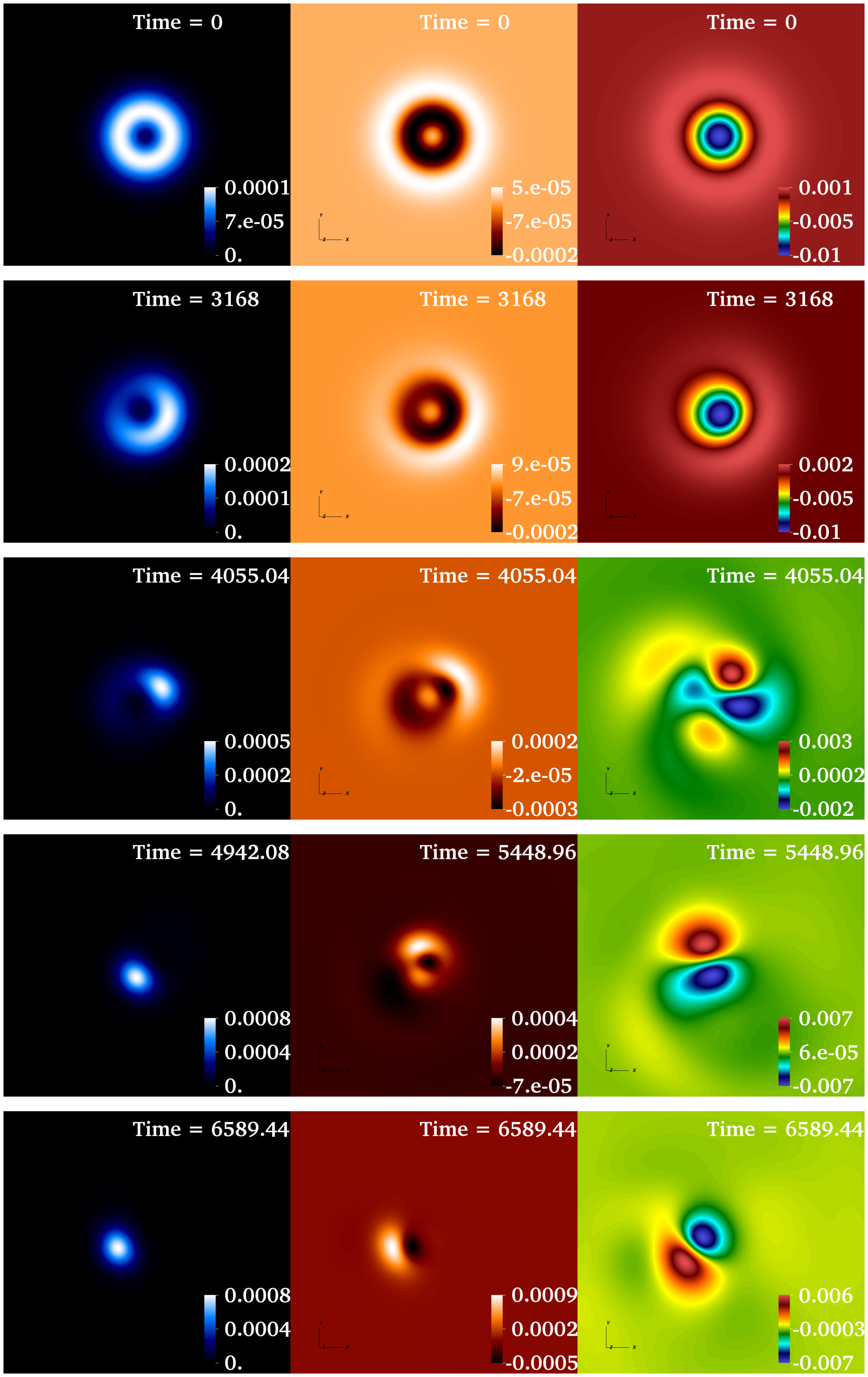}

\end{center}
\caption{ 
{\small 
{\it Left:}   
 Snapshots of the Proca Komar energy density in the $xz$-plane for hybrid ($\ell=0$ electric and $\ell=1$ magnetic) Proca star with $\omega/\mu=0.95$. {\it Middle:} Same for the angular momentum density.
{\it Right:} 
Same for the real part of the scalar potential. 
}
}
\label{evol_l1_l0}
\end{figure}

In Fig.~\ref{evol_l1_l0}, we present snapshots of the evolution of the energy density, the angular momentum density, and the real part of the scalar potential in the equatorial plane for the configuration with $\omega/\mu = 0.95$. Initially, the total Komar angular momentum is zero, although its local density is not, as illustrated in the middle panels of Fig.~\ref{evol_l1_l0}.

Similarly to the purely magnetic case, the system develops a non-axisymmetric instability dominated by a $m=1$ mode. This mode induces an axial ($m=1$) structure in the scalar potential. The final state does not correspond to a prolate Proca star; instead, the system retains part of its angular momentum distribution, leading to a rotating Proca star with a $m=1$ structure~\cite{Sanchis-Gual:2019ljs}.

The dynamics of this transition suggest a fragmentation process. This behavior is consistent with the internal structure of these configurations, where the inner core rotates in one direction while the outer layers counter-rotate. As a result, the system appears to split into a central rotating object and a secondary, smaller clump orbiting in the opposite direction, resembling a star--``moon'' configuration.

In addition, some ejection of Proca field energy is observed during the evolution, which can produce a recoil (kick) in the system. 

\subsubsection{$\ell=1$ electric + $\ell=1$ magnetic}
We now consider configurations constructed from the superposition of $\ell=1$ electric and $\ell=1$ magnetic Proca star solutions. The models analyzed are characterized by
\begin{equation}
(\omega/\mu, M\mu) = \{(0.95, 1.045), (0.93, 1.191), (0.90, 1.340), (0.85, 1.461)\}.
\end{equation}
These configurations probe a regime where both electric and magnetic components contribute non-trivially to the dynamics.

The most massive configuration, with $\omega/\mu = 0.85$, lies beyond the maximum mass supported by this family and undergoes prompt gravitational collapse, forming a black hole. The remaining models are dynamically unstable and evolve away from their initial states. In general, the evolution drives the system toward a prolate Proca star configuration, indicating that this may again represent the endpoint of the instability. An interesting exception is the model with $\omega/\mu = 0.90$, which initially relaxes toward a prolate configuration but eventually collapses to a black hole.

\begin{figure}[h!]
\begin{center}
$xz$-plane\\
 \includegraphics[height=.4\textwidth, angle =0 ]{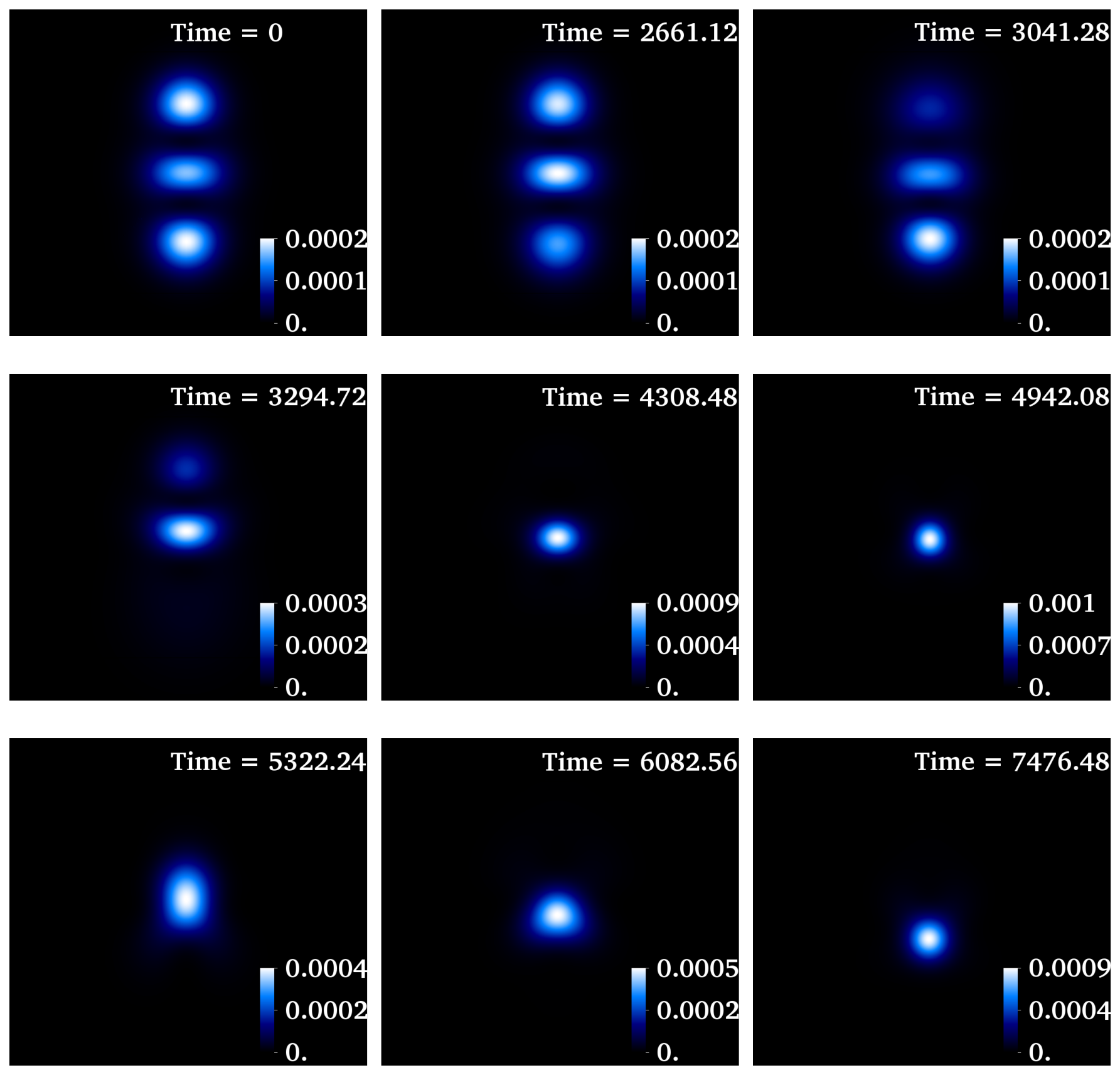}
 \includegraphics[height=.4\textwidth, angle =0 ]{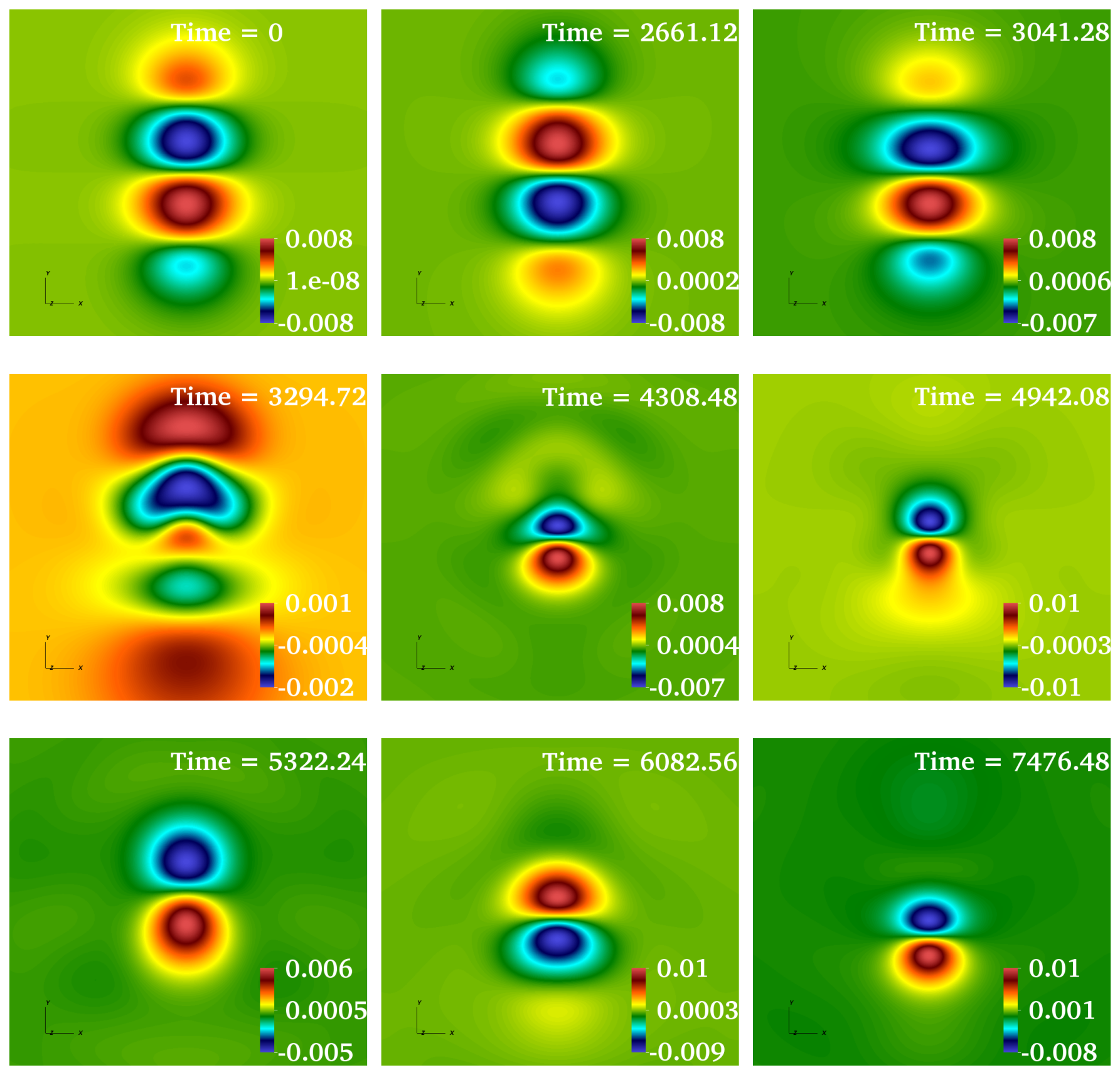}
\end{center}
\caption{ 
{\small 
{\it Left:}   
 Snapshots of the Proca Komar energy density in the $xz$-plane for hybrid ($\ell=1$ electric + $\ell=1$ magnetic) Proca star with $\omega/\mu=0.95$. Time increases from left to right and from top to bottom.
\textit{Right:} Same as in the left panels, but for the real part of the scalar potential $\mathcal{X}_{\phi}$.
}
}
\label{evol_l1_l1}
\end{figure}

In Fig.~\ref{evol_l1_l1}, we plot the evolution of the configuration with $\omega/\mu = 0.95$. The energy density, shown in the left panel on the $xz$ plane, evolves from an initial quadrupolar distribution, reflecting the superposition of electric and magnetic components, into a prolate configuration. At the same time, the scalar potential develops a  dipole as expected for a prolate Proca star. Despite these significant morphological changes, the total angular momentum remains zero throughout the evolution.

\begin{figure}[h!]
\begin{center} \includegraphics[height=.5\textwidth, angle =0 ]{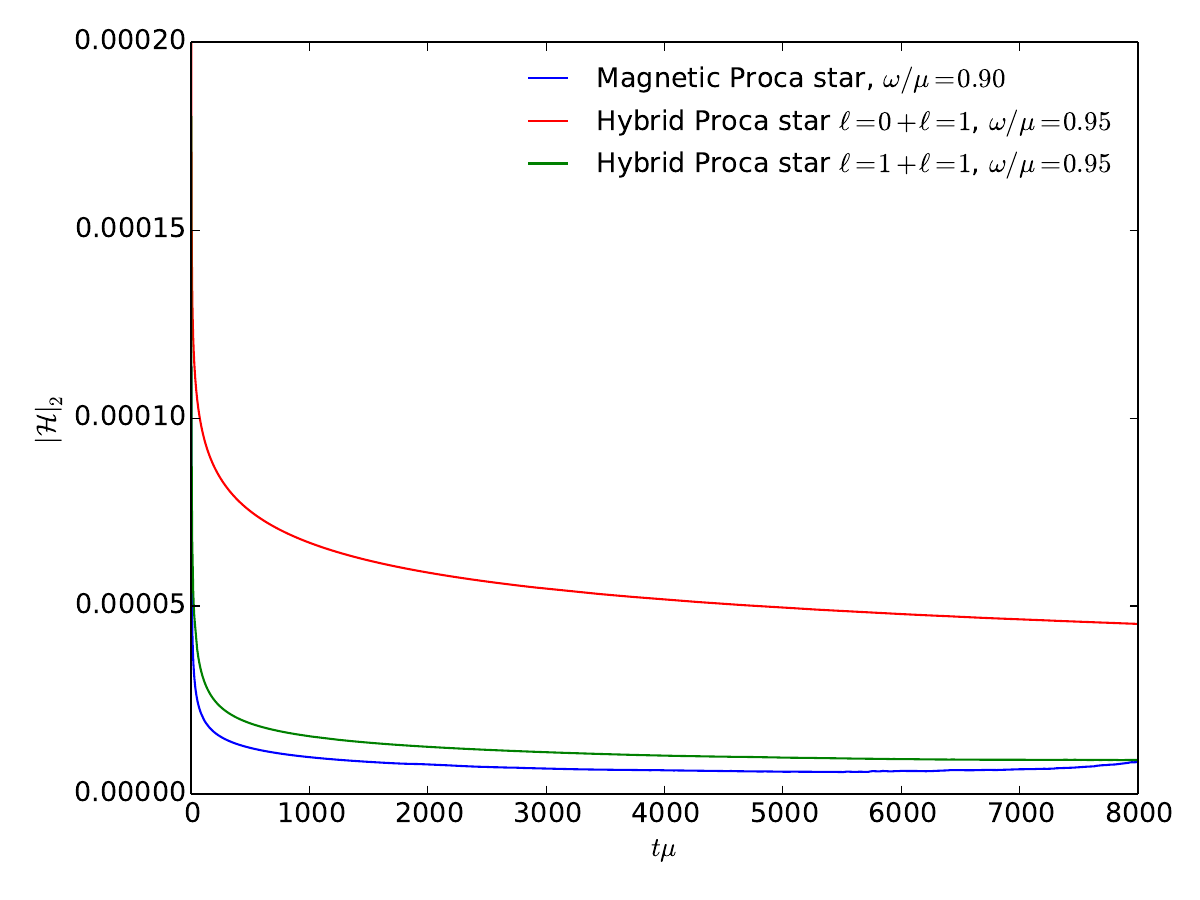}
\end{center}
\caption{ 
{\small Time evolution of the Hamiltonian constraint for all Proca star configurations considered in this section, illustrating the control of constraint violations throughout the simulations. 
}
}
\label{evol_constraints}
\end{figure}

Finally, in Fig.~\ref{evol_constraints}, we display the time evolution of the Hamiltonian constraint for the three representative multipolar Proca star configurations discussed above. The constraint violations remain well controlled throughout the evolution, with no noticeable growth during the onset and development of the instability.

\section{ Further remarks   }
\label{sec:remarks}

While solitonic solutions of the Einstein--Klein--Gordon model have been investigated for more than five decades, the exploration of analogous configurations supported by a massive complex spin--1 field remains comparatively recent. The present work contributes to this programme by revealing new sectors in the soliton landscape of the Einstein--Proca model.

Our main result is the construction of new families of static, axially symmetric Proca stars that can be interpreted as the self-gravitating counterparts of flat-spacetime Proca multipoles. These solutions naturally organize into two classes, which we dubbed \emph{electric} and \emph{magnetic}, depending on the Proca potential responsible for their support. We have analyzed their basic physical properties, including their domain of existence, mass and charge behaviour, quadrupole moments, compactness, and energy-density morphology. In addition, we have shown that these configurations admit hybrid generalizations. In particular, these hybrid solutions rotate locally while possessing vanishing total angular momentum, and in some cases possess no north-south $\mathbb{Z}_2$-symmetry.

The dynamical stability of the newly found static magnetic configurations and locally spinning hybrid solutions was investigated briefly.  Purely magnetic initial data can dynamically evolve towards electric configurations, corroborating that the electric $\ell=1$ solution reported in~\cite{Herdeiro:2023wqf}  indeed represents the true ground state among all static Proca solitons. Hybrid solutions, on the other hand, can fragment, leaving behind a spinning Proca star, with a sort of counter-rotating satellite. But other fates are possible. 

A natural extension concerns models in which the vector field acquires its mass through a Higgs mechanism. In such theories, regular Proca solitons may exist even in flat spacetime. One may therefore expect analogues of the multipolar configurations reported here to arise in that context as well. At present, only the generalizations of the electric $\ell=0$ and spinning Einstein--Proca stars are known in such models~\cite{Herdeiro:2023lze,Herdeiro:2024pmv,Adam:2024zqr,Brito:2024biy}.

Another interesting question concerns possible black hole generalizations of the solutions constructed in this work. In the absence of rotation it seems unlikely that an event horizon can be consistently inserted at the center of the static configurations. On the other hand, rotating black hole counterparts of the standard spinning solitons should exist, by analogy with the construction of Kerr black holes with Proca hair in~\cite{Herdeiro:2016tmi}. It would also be interesting to investigate whether the $J=0$ spinning configurations admit black hole generalizations, despite the absence of global rotation in the solitonic limit.

Finally, the solutions presented here are likely only the first members of a broader class of multipolar Einstein--Proca solitons. By analogy with the scalar case discussed in~\cite{Herdeiro:2020kvf}, one may expect the existence of multi-center configurations that are static but not axially symmetric. In flat spacetime the Proca multipoles admit a general expansion in real spherical harmonics $Y_{\ell m}(\theta,\varphi)$. Constructing their nonlinear gravitational counterparts with $m\neq0$ would therefore be a natural extension of the present work, although it will likely require substantially more demanding numerical methods.

\section*{Acknowledgements}

This work is supported by CIDMA (\url{https://ror.org/05pm2mw36}) under the
Portuguese Foundation for Science and Technology (FCT, \url{https://ror.org/00snfqn58}), 
Grants UID/04106/2025 
(\url{https://doi.org/10.54499/UID/04106/2025}) and UID/PRR/04106/2025 (\url{https://doi.org/10.54499/UID/PRR/04106/2025}), 
as well as the projects: Horizon Europe staff exchange (SE) programme HORIZON-MSCA2021-SE-01 Grant No. NewFunFiCO-101086251 and  2022.04560.PTDC (\url{https://doi.org/10.54499/2022.04560.PTDC}).  E.S.C.F. is supported by the FCT grant PRT/BD/153349/2021 (\url{https://doi.org/10.54499/PRT/BD/153349/2021}) under
the IDPASC Doctoral Program. N.S.G. acknowledges support from the Spanish Ministry of Science, Innovation, and Universities via the Ram\'on y Cajal programme (grant RYC2022-037424-I), funded by  MICIU/AEI/10.13039/501100011033 and by ESF+. This work is further supported by the Spanish Agencia Estatal de Investigaci\'on (Grant PID2024-159689NB-C21) funded by  MICIU/AEI/10.13039/501100011033 and ERDF A way of making Europe. The authors acknowledge computer resources provided by the Red Espa\~nola de Supercomputaci\'on (Tirant, MareNostrum5, Altamira, and Storage5) and the technical support from the IT departments of the Universitat de Val\`encia and the Barcelona Supercomputing Center (Projects No.~RES-FI-2024-2-0012 and No.~RES-FI-2024-3-0007) and by Instituto de Física de Cantabria  (IFCA) (Altamira) through Project No.~FI-2025-1-0011). Computational resources were also provided via FCT through project 2025.09498.CPCA.A3.

\bibliographystyle{hhieeetr.bst}
\bibliography{biblio.bib}

\end{document}